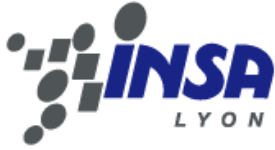
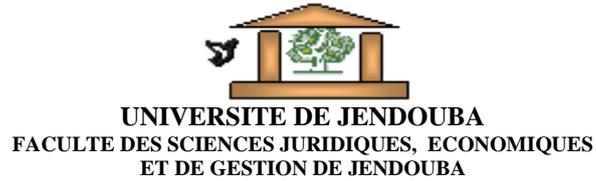
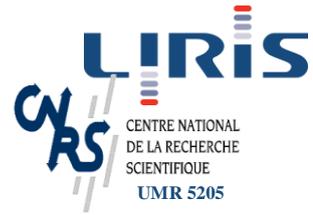

**UNIVERSITE DE JENDOUBA**
**FACULTE DES SCIENCES JURIDIQUES, ECONOMIQUES**
**ET DE GESTION DE JENDOUBA**

# MEMOIRE

**Présenté devant**

L'Institut National des Sciences Appliquées de Lyon
(Ecole Doctorale Informatique et Information pour la Société)

**Pour obtenir**

**Le Grade de Master**
Spécialité : Informatique

**Présenté Par**

**Ali Mansouri**

# Intégration des règles actives dans des documents XML

**Encadré par**

**Youssef Amghar**

**Soutenu le 27 juin 2005**

# Remerciements

*On ne peut pas passer à la page suivante sans remercier toutes les personnes qui m'ont aidé au cours de mon travail de master.*

*Je tiens à remercier :*

*Mon infinie gratitude à Mr Youssef Amghar, professeur à l'INSA de Lyon, pour m'avoir donné un encadrement efficace, sans oublier ses conseils judicieux, sa patience et le temps qu'il m'a consacré pendant toute mon master: encore merci.*

*Ma sympathie, aux professeurs Mr André Flory et Mr Robert Laurini, pour m'avoir accueilli dans le Laboratoire d'InfoRmatique en Image et Systèmes d'information (LIRIS). Merci.*

*Mes remerciements à Mr André Flory, professeur à l'INSA de Lyon, pour la lecture du rapport.*

*Mes remerciements vont à tous les membres du Laboratoire LIRIS pour leur aide de tous les instants. Merci*

*Un grand Merci à la faculté de jendouba et surtout à notre doyen Hassouna Fedhila pour ses efforts pour la reuissite de la cooperation entre l'INSA et FSJEGJ, sans oublier Mr Med salah ben yahmed et Mr mongi boulehmi qui nous ont encouragé durant le master. Encore Merci*

*À tous ceux qui ont contribué à enrichir mon travail de master, par leurs travaux, leurs idées, leurs conseils. Enfin, à tous ceux qui d'une façon ou d'une autre m'ont aidé à la réalisation de ce master*

*Un grand Merci a mon père « Taieb » qui m'a tant encouragé durant toute ma carrière scientifique.*

*À Dieu*

*À ma mère, Mon frère et mes sœurs …*

*Pour finir, un grand Merci à la personne qui m'a tant encouragé : la plus belle bénédiction de la vie : Ma fiancé Houda.*

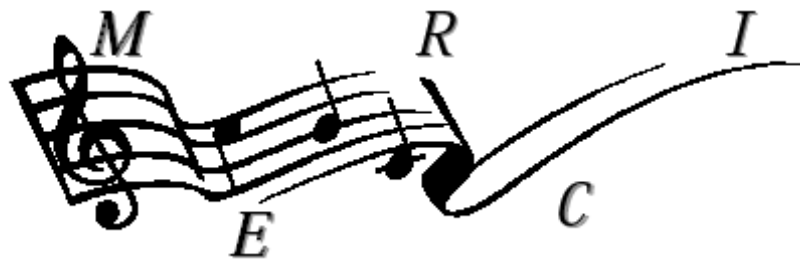

# RESUME

La gestion de la documentation technique, champ d'étude de notre travail, est une activité très intéressante pour les entreprises. En effet, les entreprises ont besoin de gérer leurs documents de la création jusqu'a l'archivage. Pour cela, le besoin d'élaboration des systèmes permettant la gestion des documents est important. En dépit des systèmes existants sur le marché (systèmes de gestion électronique des documents « GED »), on constate que ce sont des systèmes fermés qui reposent sur un modèle des données empêchant toute extension ou amélioration. De plus, ces systèmes ne permettent pas la description fine des documents et proposent des gestions de liens très peu flexibles ce qui rend la gestion de cohérence des documents très difficile.

Le LIRIS a mené des travaux pour tenter de remédier aux limites des systèmes GED. C'est ainsi qu'a été développé le système actif « SAGED ». Toutefois, SAGED s'appuie sur une approche qui sépare les règles (qui régissent la cohérence) des documents. L'inconvénient majeur de cette approche réside dans la rigidité qui existe lors de déplacements de documents d'un serveur à un autre ou d'échanges de ces mêmes documents. En effet, déplacer un document fait perdre les règles actives ECA qui lui sont associées annihilant ainsi le comportement actif du document.

Pour palier ces manques, notre contribution se base sur une approche visant à améliorer la gestion des documents en se basant sur l'introduction de règles au sein de la description des documents selon XML [BoCP01].

Dans ce contexte, nous avons proposé une couche de stockage orienté XML qui permet, grâce à une base native XML de stocker les liens, les documents et les règles dans des fichiers XML. Nous avons défini un système intelligent de gestion électronique de documents « SIGED » selon une architecture client /serveur, basée sur un composant intelligent permettant l'exécution des règles actives contenues dans un document XML et par conséquent la gestion des documents actifs.

*Mots clés* : *règles actives, XML active, introduction des services actifs, documents XML, cohérences des documents*

# ABSTRACT

The management of technical documentation is an unavoidable activity interesting for the enterprises. Indeed, the need to manage documents during all the life cycle is an important issue. For that, the need to enhance the ability of document management systems is an interesting challenge. Despite existing systems on market (electronic document management systems), they are considered as non-flexible systems which are based on data models preventing any extension or improvement. In addition, those systems do not allow a slight description of documents elements and propose an insufficient mechanisms for both links and consistency management.

LIRIS laboratory has developed research in this area and proposed an active system, termed SAGED, whose objectives is to manage link and consistency using active rules. However SAGED is based on an approach that split rules (for consistency management) and documents description. The main drawback is the rigidity of such approach which is highlighted whenever documents are moved from one server to another or during exchanges of documents.

To contribute to solve this problem, we propose to develop an approach aiming at improve the document management including consistency. This approach is based on the introduction of rules with the XML description of the documents [BoCP01].

In this context we proposed a XML-oriented storage level allowing the storing of documents and rules uniformly through a native XML database. We defined an intelligent system termed SIGED according a client/server architecture built around an intelligent component for active rules execution. These rules are extracted from XML document, compiled and executed.

# TABLE DE MATIERES

# INTRODUCTION GENERALE

Depuis les dernières années, la gestion de données distribuées change radicalement depuis l'apparition du langage XML et des services Web. En effet, XML émerge comme un langage universel pour gérer et échanger des données hétérogènes non seulement sur le Web, mais dans les systèmes d`information en général, il s`impose aussi comme un langage commun pour échanger les données sur Internet ou Intranet.

## 1. Contexte générale

La gestion des documents XML étant de plus en plus intéressante dans les entreprises, pour cela, il apparaît nécessaire de concevoir des outils ou des systèmes permettant de gérer et d'intégrer des sources de données hétérogènes en XML. En outre, la génération des systèmes de gestion des documents XML permet à un organisme de gérer efficacement tous ses documents qui sont créés et conservés sous forme électronique. Plusieurs systèmes ont été établis pour résoudre les problèmes liés à la gestion de la documentation technique. Notamment, ces systèmes permettent de sauvegarder tous les documents XML pour l'entreprise, ce qui facilite leur suivi et leur récupération. Ces systèmes peuvent également être reliés à des logiciels de gestion de contenus Web pour accélérer la publication de documents sur l'intranet ou sur le site Web d'un organisme. En bref, les systèmes de gestion des documents classiques permettent aux utilisateurs, de modifier, de supprimer, d'ajouter des documents qui sont stockés dans une base de données relationnelle.
Avec ces différents systèmes de gestion des documents XML, les serveurs Web sont passifs puisqu'ils n'interagissent pas automatiquement aux requêtes de client.

Le travail sur le besoin d'un comportement actif pour les serveurs de document, a conduit LIRIS à développer un projet SAGED [Alva003] qui vise à étendre les fonctionnalités de ces serveurs grâce a l'introduction des règles ECA. En effet, le système SAGED se base sur un module actif lui permettant la gestion des documents en utilisant des règles actives. Dans ce système, les règles sont stockées dans une base séparément des bases de documents d'où une rigidité pour la manipulation des documents. Aussi, la couche de stockage est basée sur les bases des données relationnelles ce qui nécessite un middleware pour exécuter les requêtes des utilisateurs.

## 2. Problématique :

Le système de gestion des documents « SAGED » [Alva003] s'appuie sur une approche qui sépare les règles des documents. L'inconvénient majeur réside dans la rigidité à déplacer les documents d'un serveur à un autre ou d'échanger ces mêmes documents.
En effet, deplacer un document fait perdre les règles actives ECA qui lui sont associées perdant ainsi le comportement actif du document. Par ailleurs, nous souhaitons développer une approche tout - XML aussi bien pour les documents que pour les règles.
Nous nous appuierons pour cela sur les travaux de Angela Bonifati, Stefano Ceri, Stefano Paraboschi [BoCP01], Serge Abiteoul et Omar Benjelloum [AbBM04] qui proposent d'étendre XML par l'ajout ou l'introduction des mécanismes actifs (des règles actives ECA) à travers ce qui est actuellement appelé Active XML
Aussi nous supposons de gérer les règles actives comme partie intégrante de document XML.
Le but de ce stage est donc de spécifier un système intelligent « SIGED »qui résout les problèmes de cohérence de document dans une approche tout -XML. Ce système est orienté tout XML, se base sur des langages XML prédéfinies pour exécuter les règles et pour gérer les documents.

## 3. Organisation du document :

Notre mémoire est organisée en trois chapitres, le premier est consacré pour l'étude de l'état de l'art des différents systèmes de gestion des documents, le deuxième est consacré pour l'étude du système de gestion des documents « SAGED » qui est notre système de référence et le troisième décrit notre système «SIGED» proposé pour la gestion électronique des documents.

Chapitre 1 intitulé *« l'état de l'art » :* dans ce chapitre nous présentons les différents systèmes de gestion électronique des documents .Nous présentons principalement les différentes approches pour gérer les documents (approche hyperbase, approche hypertexte).

Chapitre 2 intitulé *« système de référence SAGED »* : dans ce chapitre nous décrivons brièvement le système de gestion électroniques des documents SAGED. On décrit l'architecture de ce système avec ses différents composants. Nous décrivons aussi les différentes fonctionnalités apportées par ce système. Nous terminons ce chapitre par une analyse critique du système SAGED et nous proposons notre contribution qui tend à remédier aux limites de système SAGED.

Chapitre 3 intitulé *« système de gestion électroniques des documents « SIGED »* : ce chapitre est consacré essentiellement à la présentation de notre système « SIGED » basé sur un composant intelligent. D'abord, nous représentons la couche de stockage de système SIGED sous forme des fichiers XML (les différents composant de la couche de stockage de système « SIGED » sont présentés sous forme des fichiers XML) .Ensuite nous décrivons l'architecture de système « SIGED » et nous décrivons aussi le mécanisme de l'exécution des règles actif par le composant intelligent. Nous terminons ce chapitre par une description en UML de l'exécution de la règle active dans le système « SIGED ».

# Chapitre.1 : Etat de l'art

## 1.1 Introduction

La gestion des documents est le contrôle informatisé de documents électroniques au cours de tout leur cycle d'existence, depuis la création initiale à l'archivage final. Les documents électroniques que les systèmes de gestion les gèrent peuvent comprendre n'importe quel type numérique, images, fichiers en HTML, en SGML ou en format de documents transférables (PDF), graphiques, feuilles de calcul et documents de traitement de texte.

La gestion des documents permet aux organisations d'exercer un contrôle sur la production, le stockage, la gestion et la diffusion des documents électroniques, de réutiliser plus efficacement l'information et de mieux contrôler le déroulement des opérations relatives aux documents. Cette activité de gestion est supportée par un Système de Gestion Électroniques de Documents (Systèmes de GED), dont l'objectif est d'assurer toutes les fonctionnalités dans la chaîne documentaire : depuis leur création jusqu'à leur archivage, en passant par les étapes de révision, d'approbation et diffusion [Amgh97]

Le terme «système de gestion des documents» sous-entend une vaste *collection* de systèmes qui sont, de façon générale, reliés et qui exécutent une ou plusieurs fonctions. Il s'agit d'un type de technologie de l'information relativement nouveau qui n'est pas encore défini et que l'on utilise pour coordonner la gestion, le stockage et le repérage des documents électroniques.

Dans ce chapitre, on décrit les différents systèmes de gestion des documents existant sur le marché et on présente les différentes fonctionnalités apportées par ces systèmes.

## 1.2 Base de documents (système hyperbase)

Les systèmes hyperbases sont des systèmes hypertextes ou hypermédia permettant de stocker ou bien d'enregistrer séparément les données et les liens dans des bases de données.

### 1.2.1 Définition d'un système hyperbase

« *Un système est de type hypermédia lorsqu'il est possible de gérer des données et des liens en utilisant un système de gestion de base de données* » Nùrnberg [NLSS96], Wiill [wiil93], Schùtt et Streitz [SchS90].
Ces systèmes permettent la conservation et le stockage des documents. En effet, la gestion des hyperdocuments dans la base des données est assurée par les systèmes hyperbases en utilisant un SGBD.

### 1.2.2 Définition d'un hyperdocument

D'après Pinon, Laurini [PiLa90], un hyperdocument est définit comme suit :
« *Un hyperdocument est considéré comme un ensemble de documents ou fragments de documents appelés nœuds et reliés entre eux par des liens* »

Dans ces systèmes, c a d les systèmes hyperbases, les données sont traitées à l'intérieur de l'application, ils sont récupérés du module de stockage. Ce cas est pareil, dans les systèmes de gestion de base de données, puisque les données sont stockées et récupérés via une base de donnés.

### 1.2.3 Architecture d'un système hyperbase

L'architecture d'un système hypermédia est composée de trois couches. (Fig.1.1)
*La couche application* : cette couche regroupe un ensemble d'outils comme les éditeurs graphiques, les éditeurs de texte, des outils pour la représentation multimédia.
*La couche hyperbase* : cette couche interagit avec la couche de stockage à travers une application hypermédia. Cette application est supportée par un modèle des données [Alva03] .
*La couche stockage* : cette couche contient tous les fragments et les liens et tous les nœuds d'un document.

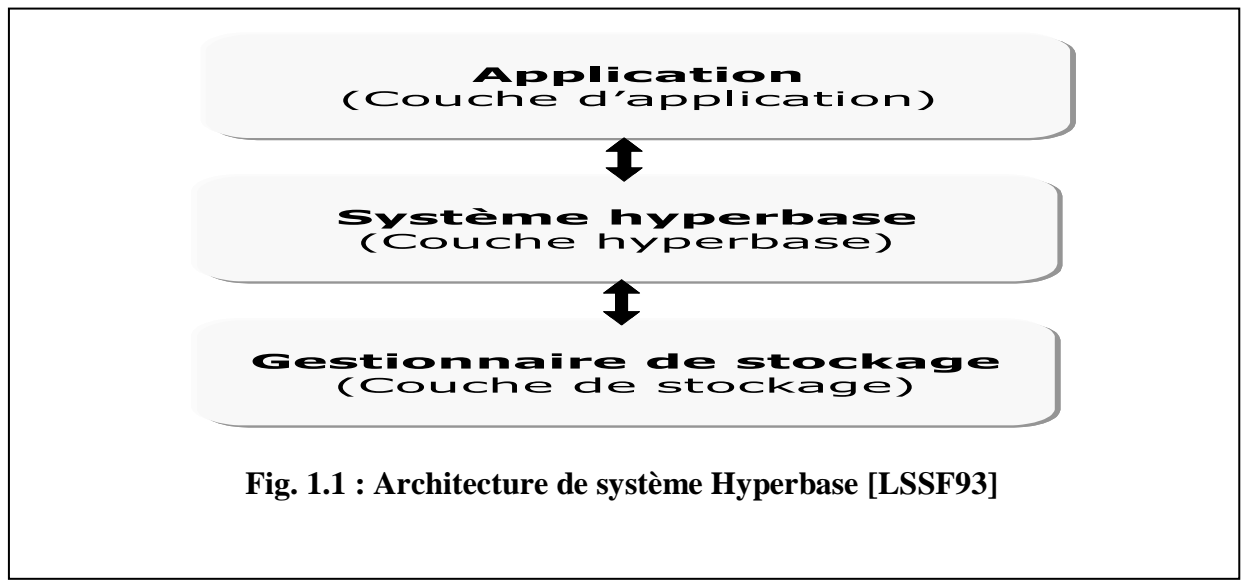

**Fig. 1.1 : Architecture de système Hyperbase [LSSF93]**

Selon Davis [Davi95], on peut caractériser les architectures des systèmes hyperbase par sept critères :

- L'échelle : concerne la taille de l'hyperbase est mesurée en octets d'information, nombre de noeuds…
- L'ouverture : c'est le degré d'interopérabilité entre des applications sans leurs imposer des restrictions par apport au modèle de données [Davi95].
- La distribution : possibilité de repartir les systèmes hyperbases sur un réseau.
- L'hétérogénéité : ce critère concerne les différents modèles des données capables de supporter les systèmes hyperbases.
- L'extensibilité : c'est la capacité des modèles des données de supporter des nouvelles opérations et abstractions.
- Le traitements des données : le système doit être capable de supporter des traitements sur les données.
- Une plate forme de développement : le modèle doit être capable de supporter la coopération entre plusieurs utilisateurs.

Ces différents critères caractérisent les différentes architectures des systèmes hyperbases ainsi que les systèmes hypermédias.

D' après Alvarez [Alva03], l'inconvénient de l'architecture des systèmes hyperbases est que l'interface entre l'application et la couche de l'Hyperbase est laissée à la charge de l'application, d'où le besoin d'utiliser généralement un modèle client serveur pour la communication.

En effet, cette architecture exige que chaque application hypertexte implémente le code pour se connecter à la couche API de l'Hyperbase. Aussi, les applications doivent être des applications d'hypertexte capables de recueillir les requêtes de l'utilisateur, demandant des informations sur des aspects tels que les caractéristiques des liens, l'affichage et la redirection vers d'autres applications.

On étudie dans la section suivante les différents services de lien utilisés pour gérer les interactions entre les utilisateurs et les systèmes.

## 1.3 Services des liens pour les systèmes hyperbases

Les services des liens permettent la gestion des interactions entre l'utilisateur et les systèmes hyperbases. C'est pour cela que ces systèmes (les systèmes hyperbases) disposent plusieurs services de liens pour pouvoir interagir avec les utilisateurs.

Parmi ces services on peut citer :

### 1.3.1 Le Modèle de Dexter

D'après LEGETT [LeSc94], ce modèle est composé de trois couches :
*La couche d'exécution* est présentée comme l'environnement de présentation, permet l'affichage hypertexte et gère l'interaction entre l'utilisateur et le système, et *la couche de stockage*, c'est au niveau du quelle se fait le stockage du contenu c a d les noeuds et des liens dans une base de données.

Ces deux couches possèdent une série de fonctions et de propriétés permettant de gérer le réseau hypertexte constitué d'éléments génériques (nœuds et liens).

La troisième couche, c'est *la couche de représentation interne* permet de présenter la structure des composants internes et définit essentiellement le contenu et la structure des informations.

Nous détaillons par la suite l'architecture du modèle Dexter selon HARDMAN Lynda, BLTERMAN C.A Kick, ROSSUM V [HaBR94].

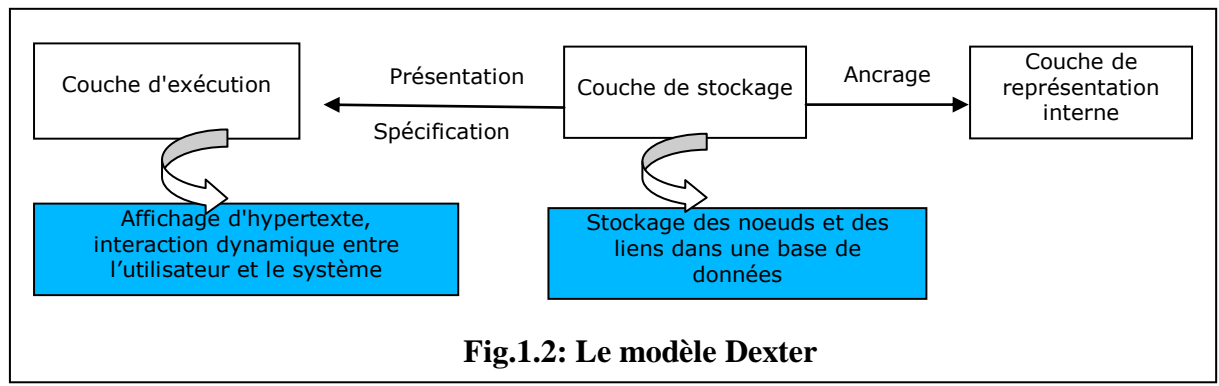

**Fig.1.2: Le modèle Dexter**

Notons que le modèle Dexter ne permet pas de définir un modèle de structure ou un type de document. En effet, il spécifie uniquement un mécanisme d'adressage ou de localisation nommé ancrage pour permettre l'accès aux contenus et aux structures des composants.

*Inconvénients du modèle Dexter :*

*1-* ne précise pas une structure interne des données, mais spécifie les données qui sont logiquement utilisables comme des composants.

*2-* présenté une rigidité au niveau du concept de granularité des données, c a d il ne permet pas une description fine des documents.

*3-*l'insuffisance de support pour des composants composites [LeSc94], ses possibilité d'évolution limitées [LeSc94], la limitation des comportements pendant la navigation [LeSc94]

*4-*le manque de support pour gérer la relation temporelle entre les données complexes (par exemple synchronisation entre l'audio et la vidéo) [HaBR94].

### 1.3.2 Le modèle Microcosm

Le modèle de Microcosm est supporté par le groupe de travail OHS (Open Hypermedia System), [OHSW98].Il a été conçu dans le but d'avoir un système sur mesure, évolutif et à la capacité d'effectuer les différents types de recherches.

FOUNTAIN M. Andrew, HALL Wendy, HEATH Ian et DAVIS Hugh décrivent Microcosm comme un système qui permet aux utilisateurs d'organiser et de naviguer dans de volumineux entrepôts de données hétérogènes [FHHH90].

Le modèle de Microcosm est composé de trois couches :

*La couche application* : gère les interactions entre le client et le service de lien

*Service de lien* : un composant actif, il a le rôle d'un middleware entre la base de donnés et la couche application

*La base des données* : sert à stocker les données, les documents, les liens …

Notons aussi que l'architecture décrite (Voir fig.1.3) permet de maintenir les liens dans une base à part, et de réaliser des opérations de traitement sur les liens. En permettant à des applications tiers d'accéder aux données de l'Hyperbase, Microcosm se décharge de la récupération des données qui composent le nœud. La couche de stockage de Microcosm est le système de fichiers même [Alva03].

Nous décrivons, ci-dessous dans la figure 1.3, l'architecture générale du Microcosm [Davi95]

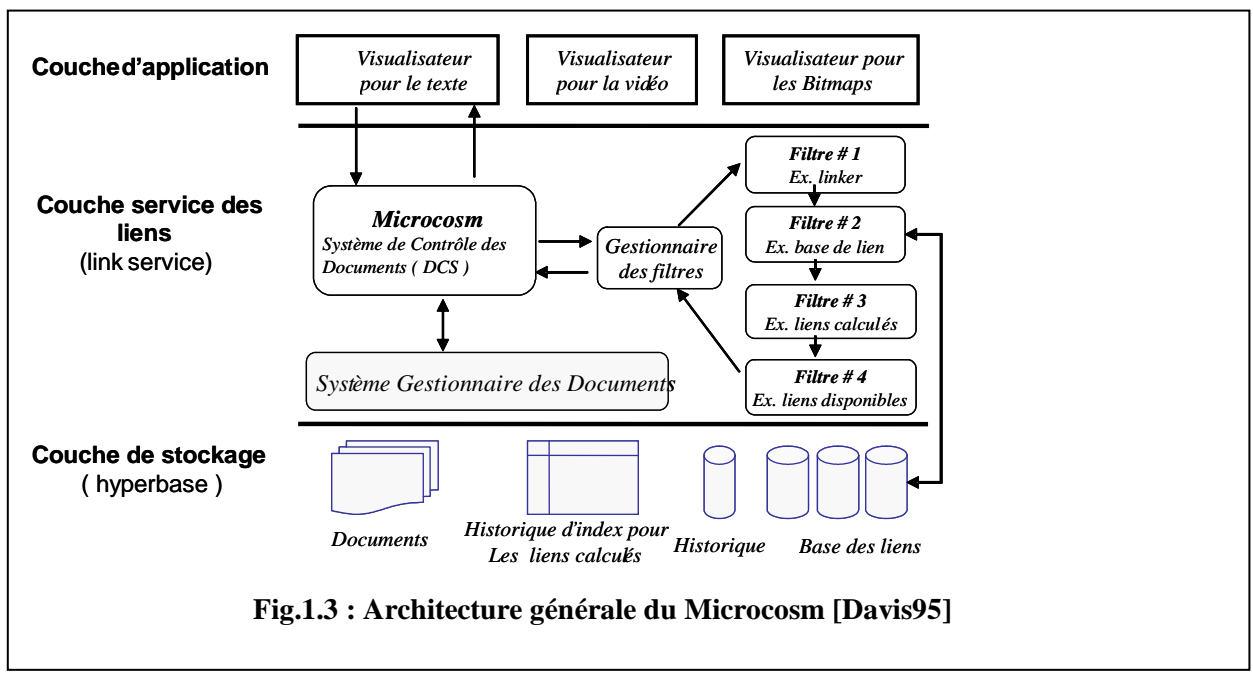

**Fig.1.3 : Architecture générale du Microcosm [Davis95]**

*Avantages:*

D'après ALVAREZ [ALva03], le système Microcosm a plusieurs avantages, dont on peut citer:
**1-** Stocke des informations sur les liens et les ancres (les ancres sont les extrémités des liens) du lien, de façon distincte des données.
**2-** Facilite la création des liens génériques.
**3-**Permet l'intégration **d**es applications externes ou « tiers » dans des systèmes hypertexte, à moindre coût.
**4-**Permet de créer des liens en lecture seule: socker les ancres du lien de façon séparées des données permet à un utilisateur de naviguer dans des données stockées sur des CDROMs, des vidéodisques, des DVDs, ou d'autres médias en lecture seule.

## 1.4 D'autres systèmes hyperbase

Il existe d'autres systèmes de gestion des hyperbases qui définissent l'architecture selon laquelle les données contenues sont stockées et par conséquent, gérées.
Parmi ces systèmes, on peut citer :

### 1.4.1 HyperWave (Hyper-G)

HyperWave est un vrai exemple de système Hyperbase intégré. Il était précédemment connu sous le nom d'Hyper-G [AnKM95].
Les liens et les données de ce système sont enregistrés dans un SGBD-OO
Dans ce système les liens sont stockés dans une base de liens qui est séparé de la base de document

### 1.4.2 GHIS –Système d'information Hypermédia Géographique

GHIS est un système d'information hypermédia géographique avec des fonctionnalités pour la gestion des liens [AsCh93]. GHIS a été développé dans le cadre du projet « *Défense Science and Technology, Organisation's Geographic Hypermédia Information System* ».
Ce système est considéré comme un navigateur de la base de données puisque toutes les données sont stockées dans un SGBD

### 1.4.3 HyperDisco

Le système HyperDisco est un système de gestion des hyperbase qui a été connu comme un OHS-HBMS (*Open Hypermédia System - Hyperbase Management System*), car il est basé sur un système de gestion Hyperbase [WiLe97].

Ce système a une architecture hybride puisqu'il se base sur deux architectures : l'architecture Hyperbase et l'architecture OHS. Le système HyperDisco permet la création de lien dans des formats arbitraires, ces liens peuvent être créés dans la base ou bien en dehors de la base, il permet aussi de stocker les données qui sont considérées comme une entrée de la base.

## 1.5 Les systèmes de gestion des documents

En plus des systèmes des gestions de bases des données comme les hyperbases que nous avons déjà présenté dans la section précédente, nous présentons aussi les systèmes de gestion des documents qui permettent le stockage et la gestion des documents dans une base des données.
Parmi ces systèmes, il existe les systèmes GED (systèmes de gestion électronique des documents) et les systèmes SAGED (systèmes actifs de gestion électronique des documents).

### 1.5.1 les systèmes de gestion électroniques des documents (GED)

Les systèmes GED assurent plusieurs fonctions pour un document. En effet, ces systèmes permettent de rechercher les documents, de suivre les modifications subites par ces documents.
De même, les systèmes GED permettent aux membres d'une entreprise de stocker et de rechercher les documents indépendamment de leurs formats que ce soit format HTML, XML ou n'importe quel autre format.

*Inconvénients* :

D'après ALVAREZ [Alva03], les systèmes GED sont des systèmes fermés car ils se reposent sur des modèles de données inextensibles :
- Ne permettent pas la gestion de la cohérence des documents.
- Ne permettent pas la description des documents avec une granularité fine c a d au niveau des noeuds si on parle d'un document sous forme arborescente
- Ne permettent pas de gérer les liens inter et intra documentaires

Le couplage entre le serveur http et les systèmes GED permet de construire un environnement X-NET dans le quel le serveur http a un comportement passif pour la maintenance de la cohérence des bases de documents.
Pour remédier aux problèmes des systèmes GED, LIRIS a développé un système de gestion électronique des documents « SAGED » qui est décrit dans la section suivante.

### 1.5.2 Le système SAGED (serveur actif de GED)

Nous pouvons considéré le système SAGED comme une évolution des systèmes GED puisqu' il offre plusieurs fonctions pour la gestion des documents durant tout leur cycle de vie.
En effet SAGED, permet le contrôle d'accès aux documents grâce à l'identité de l'auteur, et aux permissions d'accès, aussi, il permet le contrôle des documents en assurant leurs intégrités globales dans la base durant leurs cycles d'existence.
En plus le système « SAGED » a plusieurs avantages dont on peut citer :

*Les avantages de SAGED* : d'après ALVAREZ [Alava03], le système SAGED présente plusieurs avantages dont on peut citer :
 *Un système actif* : capable de répondre rapidement à des événements comme la création d'un nouveau documents, l'ajout ou la suppression d'un lien etc.
 *Système compatible avec des formats des donnés divers* : compatible avec n'importe quel format que ce soit PDF, HTML, XML etc.
 *Système modulaire* : la mise en œuvre de SAGED est indépendante des caractéristiques des systèmes d'information
 *L'indépendance du système d'information* : le système SAGED ne perturbe pas l'organisation interne des données. Ce système peut être inséré ou supprimer sans modifier l'organisation des donnés

*Inconvénients :*

Le système SAGED présente une grande évolution par apport aux systèmes GED, mais il présente, de notre point de vue, des inconvénients. En effet les règles sont stockées dans une base indépendante des bases des documents.

Ce système utilise une base de données relationnelle pour stockés les documents XML et non pas dans une base XML native pour avoir une gestion efficace des documents.

L'inconvénient majeur réside dans la rigidité à déplacer les documents d'un serveur à un autre ou d'échanger ces mêmes documents. En effet, déplacer un document fait perdre les règles actives ECA qui lui sont associées perdant ainsi le comportement actif du document.

C'est autour de ce problème qui tourne notre contribution. En effet, nous proposons de transformer la couche de stockage de système SAGED en une couche tout orientée XML et nous proposons aussi un système pour l'exécution de la règle active qui se base sur un composant intelligent .Ce nouveau système, nommé système intelligent « SIGED », basé sur un composant intelligent pour permettre une bonne gestion de la cohérence globale des documents.

Ce système permet la gestion des règles actives intégrées au début dans les documents XML donc ce système permet la gestion des documents XML actifs.

Ce système sera décrit en détail dans le chapitre 3.

## 1.6 Conclusion

Nous pouvons retenir de ce chapitre qu'il existe plusieurs approches pour la gestion des documents. Ces approches sont souvent complémentaires même leurs modèles et leurs architectures sont différentes. Nous pouvons retenir aussi que les systèmes de gestion des documents offrent des fonctionnalités permettant une bonne gestion des documents mais pas totalement efficace même le système SAGED, qui est considéré plus efficace par apport aux systèmes GED, présente des inconvénients pour la gestion de cohérence des documents (Voir chapitre suivant).

# Chapitre.2 : Le système de référence SAGED

## 2.1 Introduction

La gestion de la documentation pose beaucoup des problèmes pour les entreprises. Plusieurs systèmes ont été établis pour la résolution des problèmes de gestion de l'intégrité et de la cohérence des documents. En effet, les systèmes de gestion des documents exécutent des requêtes et des transactions lancées par les utilisateurs et les programmes d'application. Les systèmes de gestion des documents traditionnels sont passifs en ce sens que les opérations, calculs, requêtes, manipulations de données et transactions sont exécutés exclusivement à la demande explicite des utilisateurs et des programmes d'application. Par contre, les systèmes actifs ont la capacité de réagir à des événements spécifiques pour entreprendre des actions prédéfinies. Ces systèmes sont capables aussi de détecter des situations ou des faits d'intérêt et de réagir pour exécuter automatiquement certaines actions [AAmR02].

Cette capacité de réaction est réalisée par l'adjonction d'un système de règles actives Evenement-Condition-Action.

Parmi ces systèmes, nous citons le système actif de gestion électronique des documents SAGED. En effet, ce système est considéré comme un système évolué puisqu'il a la capacité de réagir à des événements spécifiques pour entreprendre des actions prédéfinies.

Ceci est réalisé, par l'ajout des services actifs pour le document XML en intégrant la notion des règles actives.

La règle active est caractérisée par le triplet suivant :(événement « E », condition « C », action « A ») La sémantique de cette syntaxe est la suivante : "Quand l'Evénement spécifié dans la définition de la règle est détecté, la Condition est évaluée, Si la Condition est vraie alors l'Action est exécutée".

Chacun de ces triplets possède un aspect bien déterminé. En effet, l'événement « E » est considéré comme le déclencheur de la règle. Une fois la règle est déclenchée, la condition « C » est évaluée, si l'évaluation est vraie c a d le résultat est non vide, alors l'action « A » est exécutée.

Le système de gestion des documents SAGED, présente beaucoup d'avantages au niveau de la gestion de la documentation et au niveau de l'aspect actif intégré dans ce système par l'introduction des règles actives (Voir après).

Dans ce chapitre, nous étudions le système de gestion des documents « SAGED » avec les différentes fonctionnalités lui permettant la gestion des bases de documents.

Aussi nous étudions la notion active de système offerte par l'introduction d'un mécanisme des règles actif lui permettant de réagir à des événements spécifiques pour exécuter des actions prédéfinies. Nous terminons ce chapitre par une analyse critique de ce système et par une présentation de notre contribution qui sera basé sur l'analyse critique de système SAGED

## 2.2 Le système actif de gestion électronique des documents « SAGED »

Le système SAGED est une évolution des systèmes GED, il offre un mécanisme actif en adoptant à son architecture la notion de règles actives d'où la notion active du système.

Ce système dispose de plusieurs fonctionnalités lui permettant de remplir les fonctions propres d'un système GED [Alva03],en plus d'un ensemble des fonctionnalités ajoutés par l'introduction de l'aspect actif basé sur les règles actifs(ECA).

### 2.2.1 Définition

Le système actif de gestion électronique de document (SAGED) est un système GED enrichi par des services actifs par l'introduction de règles actives. C'est un système actif capable de répondre rapidement à des événements comme la création d'un nouveau document, l'ajout ou la suppression d'un lien etc.

Ce système est capable aussi de gérer la structure, le contenu et les liens des documents.

Il se base sur une architecture client serveur qui sera présenté avec détail dans la section suivante.

**2.2.2 Architecture du système de gestion des documents « SAGED »**

Cette architecture est composée de trois couches :

- La couche application : c'est une interface homme machine qui permet au client d'accéder et de manipuler la base de documents à travers des requêtes.
- La couche serveur : assure la manipulation des données. Elle est considérée comme un client pour la couche stockage et comme un serveur pour les clients.
- La couche stockage : elle représente les bases de documents et assure le sauvegarde de données.

L'architecture de système SAGED est présentée dans la figure 2.1 :

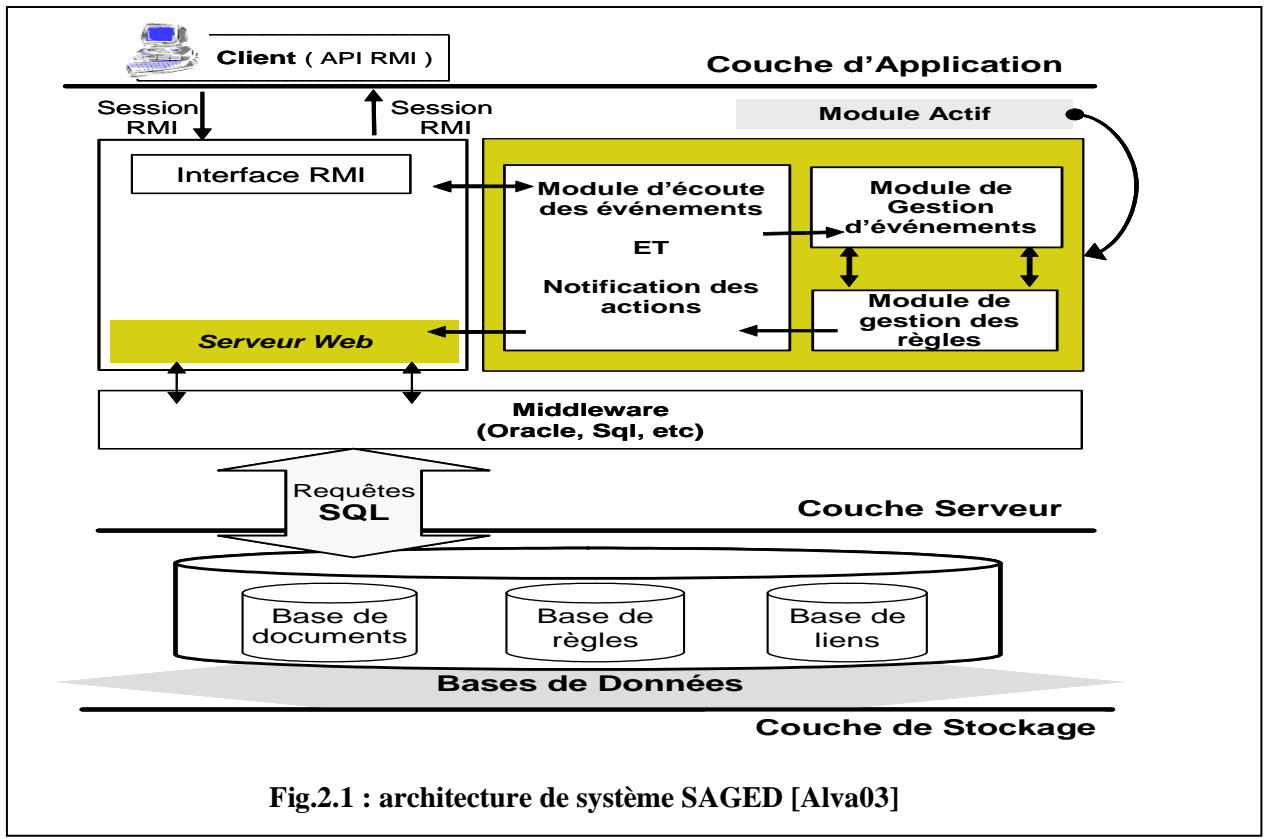

**Fig.2.1 : architecture de système SAGED [Alva03]**

Le système de gestion des documents « SAGED » présente plusieurs fonctionnalités lui permettant de gérer avec efficacité la documentation technique et l'ensemble des problèmes liés a la gestion des liens et la cohérence des documents.

**2.2.3 Les fonctionnalités de SAGED**

D'après Alvarez [Alva03], le système de gestion des documents SAGED assure plusieurs fonctionnalités dont on peut citer :

- L'accès à la base : le système SAGED XML gère les accès à la base de documents, il permet à un certain nombre d'utilisateurs d'accéder à la base en leur fournissant les droits d'accès. Ces droits peuvent être en lecture, lecture ou écriture pour apportés des modifications a la base.
- La recherche de document : SAGED permet aussi la recherche des documents selon des critères de recherche imposés par l'utilisateur (les critères de recherche peuvent être par nom, par numéro, par version, etc.).
- Une interface graphique : le système de gestion des documents SAGED offre à l'utilisateur une interface graphique lui permettant de faire des opérations de mise à jour de la base de documents depuis son poste de travail.

En plus des fonctionnalités que nous avons déjà cité, le système SAGED possède aussi un ensemble des propriétés qui le caractérisent des autres systèmes de gestion des documents.

Ces propriétés sont décrites dans la section ci-dessous.

### 2.2.4    Les propriétés de SAGED XML

D'après Alvarez [Alva03], ce système possède  aussi plusieurs propriétés dont on peut citer :

- la propriété active : la notion active du système est basée sur un module actif composé d'un module d'écoute des événements, d'un module de gestion des événements et un module de gestion de règles. A l'aide de ces modules, le système de gestion des documents SAGED peut prendre en charge les événements qui surviennent dans la base de documents. Cette propriété représente le coeur du serveur actif de documents en tenant compte des services actifs ajoutés par le module actif.
- La compatibilité avec le format de données : le système SAGED est compatible avec divers formats de données (HTML, XML, Word). Ce système est extensible puisqu'il permet la possibilité d'ajouter et de supprimer des formats de données. Ceci est à la responsabilité du module de paramétrage [Alva03].
- La propriété de modalité et de portabilité : la mise en oeuvre de système de gestion SAGED est indépendante des caractéristiques du système d'information de telle sorte qu'il ne doit pas perturber l'organisation des données. Il doit être insérer dans le système d'information, puis retirer du système d'information, sans que cela n'affecte l'organisation ou la pérennité des documents. [Alva03]

Nous avons décrit le système de gestion des documents SAGED en présentant son architecture et ses différentes fonctionnalités ainsi que ses propriétés.

Dans la section suivante, nous présentons une analyse critique de ce système et par la suite nous décrivons notre contribution qui sera  basée sur cette analyse critique.

### 2.2.5    Analyse critique du système SAGED

Malgré que le système de gestion des documents SAGED présente plusieurs avantages au niveau des fonctions offertes pour la gestion des documents, surtout au niveau de la gestion des liens et l'intégration des services actifs par l'introduction des règles actives au niveau de serveur Web pour le rendre actif au lieu de passif, il présente aussi des limites concernant la gestion de la règle. En effet les bases de règles sont séparés des bases de document d'ou on a une rigidité pour la manipulation des documents XML.

Dans le cas où la règle est intégrée dans le document XML, on a besoin d'un système qui puisse extraire et exécuter une règle.

En plus la couche de stockage du système de gestion des documents « SAGED »étant basée sur les bases de données relationnelles, nous proposons de rendre cette couche orienté XML, autrement dit le contenu des documents, les liens et les versions ainsi que les règles de cohérence suivant le modèle XML.

Aussi, le système de gestion des documents SAGED s'appuie sur une approche qui sépare les règles des documents. L'inconvénient majeur réside dans la rigidité à déplacer les documents d'un serveur à un autre ou d'échanger ces mêmes documents.

En effet, déplacer un document fait perdre les règles actives ECA qui lui sont associées perdant ainsi le comportement actif du document.

***Démarche et contribution de notre travail :***

Au vu de ce qui vient d'être énoncé et pour remédier aux problèmes déjà cités dans l'étude de l'état de l'art, et en raison des fonctions remplies par les systèmes utilisant un mécanisme de règles actives restent encore limitées, en regard des propositions émanant de la recherche, l'utilisation intensive des règles actives dans des applications réelles est très problématique en raison de leur sémantique complexe, nous proposons une solution pragmatique permettant une gestion efficace de document en offrant un aspect actif au document XML par l'intégration des règles actives dans le contenu du document.

En se basant sur les études de l'introduction des règles actives pour les services électroniques qui mettent en œuvre le paradigme ECA aux documents XML et sur l'introduction des services actifs par l'utilisation des règles actives [BoCP01] et en se basant aussi  sur l'expérience de l'introduction des règles actives dans les documents XML qui est basée sur la modélisation conceptuelle de la gestion des règles, sur la conception de base de données orientée objet active et sur des systèmes basés sur les événements [Bern04] ,nous proposons une  couche de stockage orienté tout  XML pour le cas du système intelligent de gestion électronique des documents « SIGED » que nous le décrivons dans le chapitre suivant.

Donc, en nous inspirant de :

1. L'introduction des règles actives pour les services électroniques sur le Web

2. L'introduction des schémas actifs par l'utilisation des règles actives [BoCP01]

3. Les documents XML actives [AbBM04]

4. Des recherches en bases de données orientées -objets [Bern04].

En se basant sur ces différentes approches pour la gestion des documents XML, nous proposons une approche qui vise à :

1. Obtenir une couche de stockage entièrement XML.

2. Développer les outils nécessaires pour extraire et exécuter des règles provenant de fichier de document XML

3. Proposer une nouvelle architecture SIGED autour de trois composants (ICM acquisition, ICM gestionnaire d'événement, ICM management)

## 2.3 Conclusion

Dans ce chapitre on a présenté le système de gestion électronique des documents « SAGED ».On a décrit les différentes parties de ce système avec ses différentes fonctionnalités.

Nous avons présenté une analyse de ce système sur laquelle nous nous basons pour remédier aux limites des systèmes de gestion des documents (GED) .Le système SAGED sert comme un système de référence pour notre travail.

# Chapitre.3 : Système intelligent de gestion électronique des documents (SIGED)

## 3.1 Introduction

Dans les systèmes de gestion de bases de données (SGBD) actifs, la notion de réaction automatique à des événements est offerte à travers des règles actives de la forme Evénement-Condition-Action. Ces règles sont gérées par des mécanismes spécifiques, dits actifs, intégrées dans les SGBD. Nous nous intéressons à l'introduction de ces mécanismes actifs dans les documents XML en se basant sur les études de l'introduction des paradigmes ECA aux documents XML (XML actif) [BoCP01].

D' après Serge Abiteboul, Omar Benjelloun, Tova Milo [AbBM04] un document AXML est définit comme suit :

*Active XML (AXML) est un modèle déclaratif pour la gestion de données distribuées sur le Web basé sur XML et les services Web (SOAP, WSDL). Un document AXML est un document XML pouvant contenir des appels à des services Web.*

En se basant sur les différentes approches de l'introduction des règles dans un document XML, nous proposons une approche basée sur un système qui permet d'exécuter la règle active contenue dans un document XML ,donc un système permettant la gestion des documents XML actifs. Dans cette étude, nous nous intéressons plus précisément à la définition et à la structuration d'un service de règles. Nous proposons un système intelligent adaptable « SIGED » qui permet la gestion des documents. La notion d'intelligence provient du faite de l'introduction d'un composant intelligent qui permet l'extraction des règles actives et leurs exécutions.

Dans la première partie de ce chapitre, on s'intéresse à la modélisation de bases des documents, la modélisation des bases des règles, la modélisation des bases de liens  c a d on tend a représenter ces différents bases sous forme des fichiers XML, c a d notre but est d'avoir une couche de stockage orienté tout XML.

Nous présentons, dans la deuxième partie, le système de gestion des documents « SIGED », nous présentons également son architecture  en détaillant ses différents composants. Nous présentons aussi le processus d'exécution des  règles actives par notre système SIGED.

## 3.2  Modélisation de la couche du stockage de système SIGED

Avoir une couche orientée tout XML, nécessite la modélisation de ses différents composantes (base des liens, base des documents, base des règles) sous forme XML. Nous nous intéressons en premier lieu à la modélisation de la base de règle puisque la règle active est le champ de notre travail.

Dans la section suivante, nous décrivons la règle active, nous décrivons la partie événement et ses types (primitifs, composites), la partie condition, la partie action.

### 3.2.1    Les règles actives

La notion de règle active est basée sur le triplet événement, condition, action <E, C, A>.
Elle se base sur le formalisme suivant Evénement, Condition, Action.
La sémantique générale de cette règle est le suivant : lorsqu'un événement se produit, si la condition est satisfaite, alors l'action est exécutée.
Nous traitons tout d'abord la première partie de la règle active « l'événement ».

### 3.2.1.1  L'événement

Cette partie spécifie l'événement responsable de déclenchement de la règle. C'est un changement à l'intérieur de la base. On peut distinguer deux types d'événements (voir figure 3.1)

***Événements primitifs*** : Il existe quatre catégories d'événements primitifs :
- Update : ces événements sont associés à l'opération de mise à jour. C'est une opération concernant la base des données ou la base des documents

- Add : ces événements correspondent aux opérations d'ajout d'un document ou d'un fragment des documents dans la base des documents.
- Remove : ce sont des événements concernant les fonctions permettant de déplacer des documents ou des fragments des documents de la base vers un autre endroit dans la base.
- Delete : ces types d'événements concernent les opérations de la suppression des documents ou des fragments de la base de documents.

Tous les événements déjà cités sont des manipulations d'entités c a d sont associés aux manipulations de la base : créations et destructions d'objets, modifications d'objets et de valeurs, insertions et suppressions dans des collections, appels des méthodes.

***Événements composites*** : ces événements sont construits récursivement à partir d'événements primitifs ou par des événements qui sont eux-mêmes des événements composites connectés par des opérateurs. Ces opérateurs peuvent êtres des conjonctions, des disjonctions, négation…

- L'opérateur disjonction « E1 OR E2» : permet la sélection d'une règle si on détecte l'événement E1 ou l'événement E2
- L'opérateur conjonction « E1 AND E2» Fait la sélection des règles lorsque les deux événements E1 et E2 sont détectés
- L'opérateur séquence « E1 : E2 » : appelés aussi, des événements logiques, l'événement E1 se déclenche après l'événement E2

***Autres types d'événements***

Il existe aussi d'autres types d'événements comme les événements transactionnels qui sont associés au début et à la fin de l'exécution des transactions. On trouve aussi les événements temporels: ces événements sont liés au temps et générés par le système.

Dans notre étude on s'intéresse aux événements suivants : Update,Add, Remove, Delete

L'ensemble des événements que ce soit primitif ou composés peut être représenté sous forme de la hiérarchie suivante :

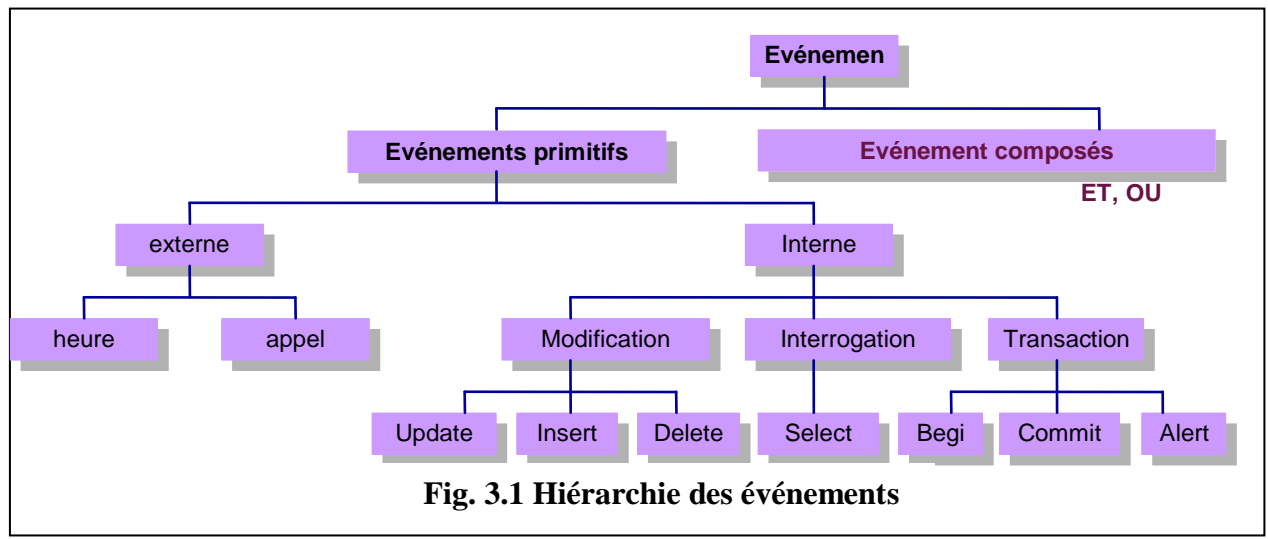

**Fig. 3.1 Hiérarchie des événements**

### 3.2.1.2 La condition

La condition est une expression qui retourne une valeur booléenne lors de l'évaluation en vue d'exécuter une action. Elle est satisfaite si la valeur retournée est non vide ou vraie .Cette partie de la règle active peut être exprimée par une conjonction des requêtes. La condition est vraie si toutes les requêtes retournent une réponse non vide.

### 3.2.1.2 L'action

La partie action de la règle active peut être un programme pour modifier la base des documents. Elle est exécutée si la condition est satisfaite. L'action est un ensemble d'opérations à exécuter qui peut être des appels des méthodes ou des fonctions.

L'exécution de l'action peut être l'origine de déclenchement d'autres événements.

La section suivante est consacrée pour la modélisation de la règle active ainsi que la modélisation de ses différents composants.

### 3.2.2 La modélisation de la règle active

Le modèle de règle permet d'exprimer l'activité de cette règle. En effet, les règles sont des instances d'une classe unique de règle. La règle active est composée par les attributs suivants à savoir l'événement, la condition, l'action. Donc la modélisation des règles se fait par la modélisation de ses différents composants.

#### 3.2.2.1 Modélisation de l'événement

On présente l'événement correspondant à la règle sous forme XML. On utilise une DTD (document type définition) pour exprimer l'événement sous forme des balises XML.
On décrit ci-dessous la DTD correspondante à l'événement d'une règle :

```
< ! ELEMENT événement (#PCDATA)>
< ! ELEMENT événement (événement primitif, événement composite) >
< ! ELEMENT événement primitif (#PCDATA)>
< ! ELEMENT événement composite (#PCDATA)>
< ! ELEMENT update (#PCDATA)? >
< ! ELEMENT remove (#PCDATA)? >
< ! ELEMENT delete (#PCDATA) ?>
< ! ÉLEMENT add (#PCDATA)?>
< ! ÉLEMENT événement composite (événement primitif, (opérateur, événement primitif) +>
< ! ELEMENT événement primitif (#PCDATA)>
< ! ÉLEMENT opérateur (OR, AND)
< ! ÉLEMENT OR (#PCDATA)>
< ! ÉLEMENT AND (#PCDATA)>
```

*Exemple :*
Soit l'événement de suppression d'un document de la base des documents, on donne ci dessous une description XML de la DTD associée.

```
< ! ÉLEMENT événement composite (delete, (AND, add) +)
< ! ÉLEMENT delete (#PCDATA)>
< ! ÉLEMENT add (#PCDATA)?>
< ! ÉLEMENT  AND (#PCDATA)
```

On donne ci-dessous une description XML de la DTD associée au second élément de la règle active (la condition)

#### 3.2.2.2 Modélisation de la condition

La condition est caractérisée par un identifiant (id condition) et une description (desc condition).La représentation XML de la condition est présentée par le document XML ci dessous:

```
< ! ÉLEMENT condition +>
< ! ÉLEMENT condition (id_condition, desc_condition)>
< ! ÉLEMENT id_condition, desc_condition (#PCDATA)>
```

La troisième partie de la règle active est l'action. On donne dans la section suivante la modélisation du l'action.

#### 3.2.2.3 Modélisation de l'action

L'action est vue comme plusieurs opérations à exécuter. En effet, modéliser l'action revient à modéliser ses attributs (les opérations à exécuter).

```
< ! ÉLEMENT action+ >
< ! ÉLEMENT update (#PCDATA) ?>
< ! ÉLEMENT remove (#PCDATA)? >
< ! ÉLEMENT delete (#PCDATA)? >
< ! ÉLEMENT add (#PCDATA) ?>
```

Nous remarquons, que les actions et les événements ont la même représentation XML puisque les actions sont la réalisation des événements déclanchés au début.

### 3.2.2.4 Modélisation de la règle active

Après la modélisation des différents composants de la règle active, on peut donc la modéliser de la façon décrite ci-dessous.

- **Schéma de la règle active**

La règle, comme nous l'avons présenté ci dessus, est composée de trois composantes. Ce triplet est l'événement, la condition, l'action. Nous la représentons par le digramme de classe suivant.

La règle peut être schématisé par la figure présentée ci-dessous (voir Fig3.2):

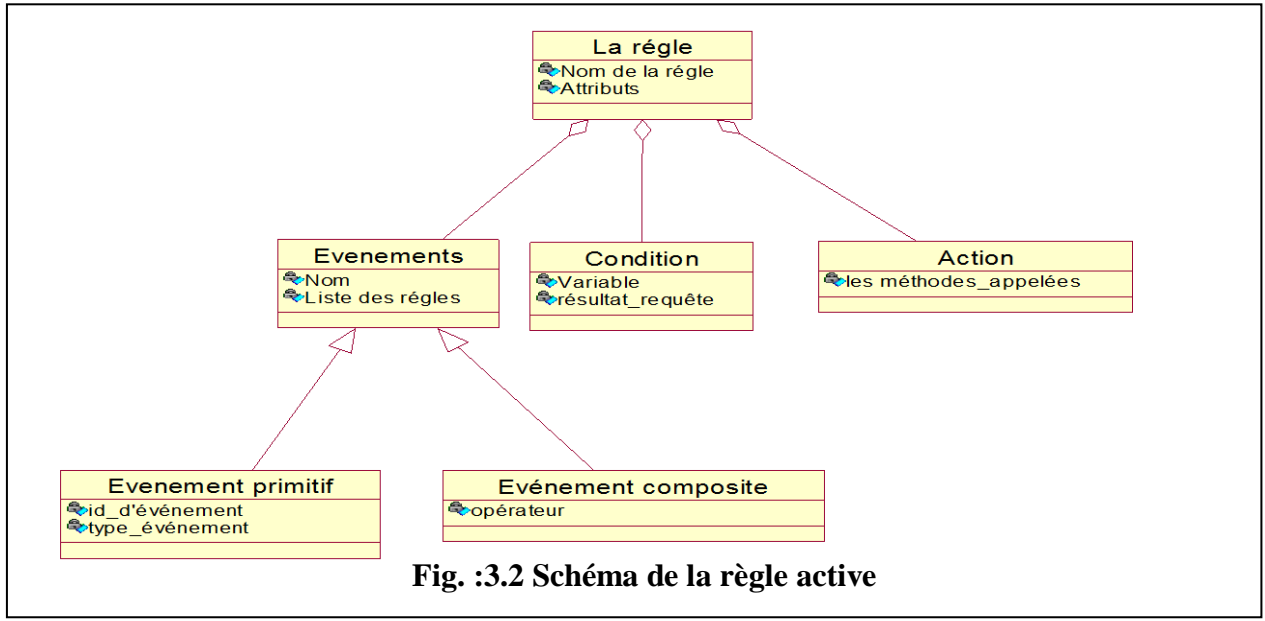

**Fig. :3.2 Schéma de la règle active**

- **La règle active sous forme XML**

La représentation de la règle active sous forme XML correspond à la concaténation des schémas XML de ces différents composants. Donc, le schéma XML de la règle active peut être représenter de la façon suivante :

```
< ! ELEMENT REGLE (événement +, condition +, action+)
< ! ELEMENT événement (événement primitif, événement composite) >
< ! ELEMENT événement primitif (#PCDATA)>
< ! ÉLÉMENT événement composite (#PCDATA)>
< ! ELEMENT updated (#PCDATA)?>
< ! ELEMENT removed (#PCDATA)?>
< ! ELEMENT delete (#PCDATA) ?>
< ! ÉLÉMENT add (#PCDATA) ?>
< ! ÉLÉMENT événement composite (événement primitif, (opérateur, événement primitif)+>
< ! ELEMENT événement primitif (#PCDATA)>
< ! ÉLÉMENT opérateur (OR, AND)
< ! ÉLÉMENT OR (#PCDATA)>
< ! ÉLÉMENT AND (#PCDATA)>
< ! ÉLÉMENT condition +>
< ! ÉLÉMENT condition (#PCDATA)>
< ! ÉLÉMENT action+ >
< ! ÉLÉMENT update (#PCDATA) ?>
< ! ÉLÉMENT remove (#PCDATA)?>
< ! ELEMENT delete (#PCDATA)?>
< ! ÉLÉMENT add (#PCDATA) ?>
```

La modélisation de l'ensemble des règles par schéma actif permet de décrire l'activité de schéma et de la rendre applicatif. On désigne par schéma actif la modélisation des événements, de la condition, l'action qui permet de rendre les différents objets de schéma actif.

L'instance de la règle active est présentée de la façon suivante :

```
<Règle>
<Événement>    un événement (primitif ou composite)
</Événement>
<Condition > une variable ou un résultat d'une requête
</Condition>
<Action> appel des méthodes
</Action>
</Règle>
```

**Exemple :**

Règle "suppression fragment document "

On update document X

 If fragment document X -> « Paragraphe »

Do {Supprimer (fragment document X)

La DTD de cette règle est donnée ci-dessous :

```
< ! ELEMENT REGLE (événement, condition, action)
< ! ELEMENT événement primitif (#PCDATA)>
< ! ELEMENT on update (#PCDATA) >
< ! ÉLÉMENT condition >
< ! ÉLÉMENT fragment X -> paragraphe (#PCDATA)>
< ! ÉLÉMENT action >
< ! ÉLÉMENT delete (#PCDATA) ?>
```

Après la modélisation des règles actives, nous décrivons la modélisation des bases de règles dans la section suivante.

### 3.2.2.5 Modélisation des bases de règles

La base des règles est un ensemble des règles qui sont associés a des documents définis dans la base des documents pour assurer leurs modifications, leurs mises a jour …

Donc la modélisation de la base des règles est vue comme la modélisation d'un ensemble des règles d'où la base des règles est vue comme un fichier XML qui contient toutes les règles sous forme des DTD.

La représentation d'un fichier XML des  règles constitué d'un ensemble des règles est  décrite ci dessous :

```
< ! ELEMENT  fichier règle (REGLE +)
< ! ELEMENT REGLE +>
< ! ELEMENT REGLE (événement +, condition +, action+)
< ! ELEMENT événement (événement primitif, événement composite) >
< ! ELEMENT événement primitif (#PCDATA)
< ! ÉLÉMENT événement composite (#PCDATA)>
< ! ELEMENT updated (#PCDATA)?>
< ! ELEMENT removed (#PCDATA)?>
< ! ÉLÉMENT delete (#PCDATA) ?>
< ! ÉLÉMENT add (#PCDATA) ?>
< ! ÉLÉMENT événement composite (événement primitif, (opérateur, événement primitif)₊>
< ! ELEMENT événement primitif (#PCDATA)>
< ! ÉLÉMENT opérateur (OR, AND)
< ! ÉLÉMENT OR (#PCDATA)>
< ! ÉLÉMENT AND (#PCDATA)>
< ! ÉLÉMENT condition +>
< ! ÉLÉMENT condition (#PCDATA)>
< ! ÉLÉMENT action+ >
< ! ÉLÉMENT update (#PCDATA) ?>
< ! ÉLÉMENT remove (#PCDATA)?>
< ! ELEMENT delete (#PCDATA)?>
< ! ÉLÉMENT add (#PCDATA) ?>
```

Après la modélisation des bases de règles nous nous intéressons dans la section suivante à la modélisation de la  base des liens qui fait partie de la couche de stockage.

### 3.2.3  Modélisation des bases des liens

La base des liens peut être aussi modéliser sous forme d'un document XML qui est présenté ci dessous. Les liens sont stockés dans un fichier XML sous forme des DTD.

Un lien est représenté de la façon décrite ci dessous :

```
< ! ELEMENT liens (liens sortants+>
< ! ELEMENT lien sortants (origine, destination+)>
< ! ELEMENT origine  (nœud)>
```

```
< ! ELEMENT nœud (#PCDATA)
< ! ELEMENT destination (nœud +)>
< ! ELEMENT nœud (#PCDATA)
```

Une instance d'un lien est présentée da la façon suivante:

```
<Liens >
<Liens sortant >
< Origine Href = ……………… …………..>
</Origine >
<destination Href =………………………>
</Destination >
</Lien sortant>
</Liens >
```

Le fichier de liens sera présenté de la façon suivante :

```
< ! ELEMENT fichier _lien (liens +)
< ! ELEMENT liens (liens sortants+>
< ! ELEMENT lien sortants (origine, destination+)>
< ! ELEMENT origine  (nœud)>
< ! ELEMENT nœud (#PCDATA)
< ! ELEMENT destination (nœud +)>
< ! ELEMENT nœud (#PCDATA)
```

### 3.2.4   Modélisation des bases des documents

La base des documents composée d'un ensemble des documents XML. Chaque document XML est caractérisé par un nom, et un ensemble d'attributs.

Dans notre approche les documents sont stockés dans un fichier XML. Donc un fichier de documents peut être décrit  sous forme d'une  DTD de la façon suivante :

```
< ! ÉLEMENT fichier doc. (DOCXML+)
< ! ÉLEMENT DOCXML (nom, définition attribut) +>
< ! ATTLIST DOCXML
ID (CDATA) #REQUIRED>
< ! ÉLEMENT nom (#PCDATA)>
< ! ÉLEMENT définition attribut (date de création, numéro de version, auteur)>
< ! ÉLEMENT date de création (#PCDATA)>
< ! ELEMENT version (#PCDATA)>
< ! ELEMENT auteur (#PCDATA)>
```

Donc en utilisant des documents XML (les DTD), nous pouvons représenter la base des liens, les bases des règles et les documents XML sous forme des fichiers XML. Nous obtenons ainsi une couche de stockage orientée XML.

Après tous ces modifications l'ensemble de fichiers des documents XML, fichiers de liens, fichier des règles sont stockés dans une base XML native puisque ce genre de base offre plusieurs fonctionnalités et plusieurs langages permettant d'offrir une grande flexibilité pour la gestion des documents .

### 3.2.5   Les bases de données XML natives

Les bases de données XML natives sont des bases conçues spécialement pour stocker des documents XML. Comme toutes les autres bases, elles possèdent des fonctionnalités telles que les transactions, la sécurité, les accès multi- utilisateurs, un ensemble d'APIs (Application Programming Interfaces) , des langages de requête, etc. La seule différence par rapport aux autres bases, c'est qu'elles sont basées sur XML, et pas sur autre chose comme dans le cas des bases relationnelles. [Rona03]. Ces bases de données XML natives sont utiles pour le stockage des contenus orientés document.

***Définition :***
Selon RONALD [Rona03], une base XML native est définie comme suit :
« Une base de données XML native définit un modèle (logique) de document XML  et stocke et retrouve les documents en fonction de ce modèle. Le modèle doit au minimum inclure les éléments, les attributs, les PCDATA et l'ordre interne du document. Quelques exemples de tels modèles sont : le modèle de données de XPath, le glossaire XML Infoset, et les modèles implicites de DOM et des événements de SAX 1.0. »

Selon RONALD [Rona03], la base XML native offre plusieurs caractéristiques permettant une grande efficacité pour la gestion des documents XML et l'application des différents traitements sur ces documents.

Parmi ces caractéristiques on peut citer :

- *Les index* : les bases XML natives supportent l'indexation des valeurs des éléments et des attributs. Les index sont utilisés pour accélérer les recherches, comme pour les bases non XML.

- *Les langages de requête* : les bases XML natives supportent des langages de requête comme XPath qui est une syntaxe (non XML) pour désigner une portion d'un document XML et comme XQuery qui est un langage de requête permettant d'extraire des informations d'un document XML

- *Les APIs [Application Programming Interfaces]* : les bases XML natives proposent des APIs. C'est une interface enrichie avec des méthodes permettant la connexion à la base, l'exploration des métas données, l'exécution des requêtes et la recherche des résultats. Ces résultats sont renvoyés sous la forme d'une chaîne XML, d'un arbre DOM (document object Model), ou bien encore d'un analyseur SAX (simple API for XML) sur le document retourné.

- *L'intégrité référentielle* : l'intégrité référentielle dans les bases XML natives peut se décomposer en deux catégories : l'intégrité des pointeurs internes (les pointeurs vers le document lui-même) et l'intégrité des pointeurs externes (les pointeurs entre documents).

A l'aide de ces différentes caractéristiques déjà cité, nous pouvons donc appliquer des différents traitements a l'aide du système SIGED basé sur un composant intelligent qui sera décrit dans la section suivante et dont le couche de stockage de ce système est basé sur les bases XML native

Dans la deuxième partie, nous représentons le système de gestion des documents « SIGED » basé sur le composant intelligent « ICM » .Nous représentons l'architecture du système avec ses différents composants. Nous décrivons aussi le fonctionnement de ce système pour l'exécution des règles actives. Nous terminons par une représentation UML de fonctionnement de système et par un exemple pour montrer l'exécution des règles actives dans notre système « SIGED »

## 3.3 Le système intelligent de gestion des documents (SIGED)
### 3.3.1 Introduction

L'objectif de notre travail est de proposer un système qui permet de détecter et de réagir aux événements provenant de l'utilisateur lorsqu'il exécute des actions de manipulation de la base des documents et qui permet aussi l'exécution d'une règle active intégrée dans un fichier XML envoyé par l'utilisateur.

Nous proposons notre système intelligent de gestion électronique des documents (SIGED) pour remédier en quelque sorte aux limites de système de gestion « SAGED ». En effet, notre système de gestion des documents « SIGED » doit permettre l'extraction et l'exécution des règles actives. Il doit être aussi capable d'interagir avec le client à travers des logiciels clients et avec la couche stockage en utilisant des langages XML comme XQuery, XPointer, XLink.

Nous détaillons par la suite les fonctionnalités de notre système, puis nous présentons son architecture et ses différents composants en détail.

### 3.3.2 Présentation de système intelligent (SIGED)
#### 3.3.2.1 Définition

Le système intelligent de gestion électronique des documents (SIGED) est un système qui comporte la notion des services actifs par l'introduction des règles actives (ECA). Il est capable de détecter des situations et de réagir avec et sans l'intervention de l'utilisateur en exécutant des actions ou des programmes. La notion ou la propriété active de système est basée sur les règles actives.

Ces règles sont gérées par un mécanisme intelligent appelé « composant intelligent » d'où on a la notion de réaction automatique de système envers des manipulations de la base des documents.

### 3.3.2.2 Les fonctionnalités de système intelligent (SIGED)

Le système de gestion de documents SIGED offre plusieurs fonctionnalités pour la gestion des documents XML. En effet, ce système doit permettre la recherche des documents selon des critères décrits par l'utilisateur , doit permettre aussi le stockage des documents , la manipulation des documents par l'exécution des opérations comme la suppression ,la mise à jour ,la création des documents…

Le système intelligent (SIGED) permet aussi la réalisation d'autres fonctions comme par exemple :

- *La gestion de l'accès a la base :*

Notre système permet la gestion de l'accès aux documents en donnant à certains utilisateurs les droits d'accéder à la base. Ce système fournit plusieurs modes d'accès aux documents comme mode de lecture seule des documents, mode de modification des documents ….

- *La fonction de recherche :*

Le système intelligent SIGED permet aussi de rechercher les documents. Les recherches des documents se fait suivant plusieurs critères imposés par l'utilisateur. Nous pouvons aussi utiliser les index fournis par les bases XML natives sur les quelles se base notre système.

Les recherches des documents peuvent êtres par nom, par date création, par auteur, ….

- *La gestion des documents :*

Le système SIGED permet de gérer les documents dans la base d'une manière efficace à l'aide des services actifs obtenus par les règles actives intégrées dans les documents XML.

- *L'interaction avec les clients :*

Le système intelligent (SIGED) interagit avec les clients en exécutant leurs règles. En effet ce système est capable d'extraire la règle intégrée dans le fichier XML émise par le client.

Aussi, il assure l'exécution de cette règle en se basant sur un composant intelligent que nous le décrivons par la suite.

- *L'interaction avec les bases des documents :*

Suite à des règles intégrées dans un fichier XML, émises par le client , le système intelligent  SIGED effectue les changements apportés par la règle sur la base des documents en utilisant des langages XML comme XQuery, XPath, XLink…

Après la définition des différentes fonctionnalités de système SIGED, on étudie dans la section suivante les composants de système intelligent  de gestion électronique des documents (SIGED)

### 3.3.2.3   Les composants du système SIGED

Le système SIGED est un système de gestion électronique des documents qui comporte des services actifs suite à l'intégration des règles actives (ECA) dans les documents XML.

Ces règles sont gérées par un mécanisme intelligent appelé « composant intelligent ».

Il est considéré comme un client pour les bases de documents et comme un serveur aux utilisateurs.

Ce système s'appuie sur une architecture client /serveur. L'architecture du système SIGED   est représentée dans la figure 3.2.

Cette architecture  est composée de trois couches :

- La couche application : appelée aussi la couche client, c'est une interface homme machine auprès du quelle le client effectue des différentes opérations sur la base en envoyant des requêtes ou des règles au couche serveur qui s'occupe de la réalisation de ces requêtes ou ces règles. Une interface RMI est utilisée pour la connexion avec les clients.
- La couche SIGED XML : c'est un sorte d'intermédiaire entre la couche client et la couche  de stockage, elle permet au client d'accéder à la base de documents et assure la manipulation des données en utilisant un médiateur XML : STYX
  - Un médiateur XML : STYX   est un modèle de médiation qui permet de décrire des ressources XML suite à l'intégration d'une ontologie composée de concepts et de rôles [ABFS02]. Le médiateur utilise des règles de traduction et un algorithme de réécriture de requêtes pour traduire les  requêtes arrivés de la part de composant intelligent  en des requêtes XPath [ClDe99] ou XQuery [ChFR01].
- La couche stockage : appelée aussi la base de donnée et assure le stockage ou la conservation des données.

### 3.3.3    Architecture du système SIGED

L'architecture de notre système se compose de  trois couches à savoir la couche application, la couche serveur, la couche stockage.

Ce système se base sur la notion client/serveur pour permettre l'interaction avec les clients comme étant un serveur et d'interagir avec la couche stockage comme étant un client.

L'architecture de notre système de gestion électronique des documents « SIGED » est basée sur les techniques suivantes :

RMI : interface pour les connexions avec les clients

STYX : un médiateur XML pour accéder aux bases des documents [ABFS02].

Xpath, Xlink et XQuery comme des langages pour accéder et extraire les documents

A l'aide de ces différentes techniques, nous présentons l'architecture de notre système de gestion des documents « SIGED ».

L'architecture de ce  système de gestion des documents « SIGED » est présentée dans la figure ci-dessous :

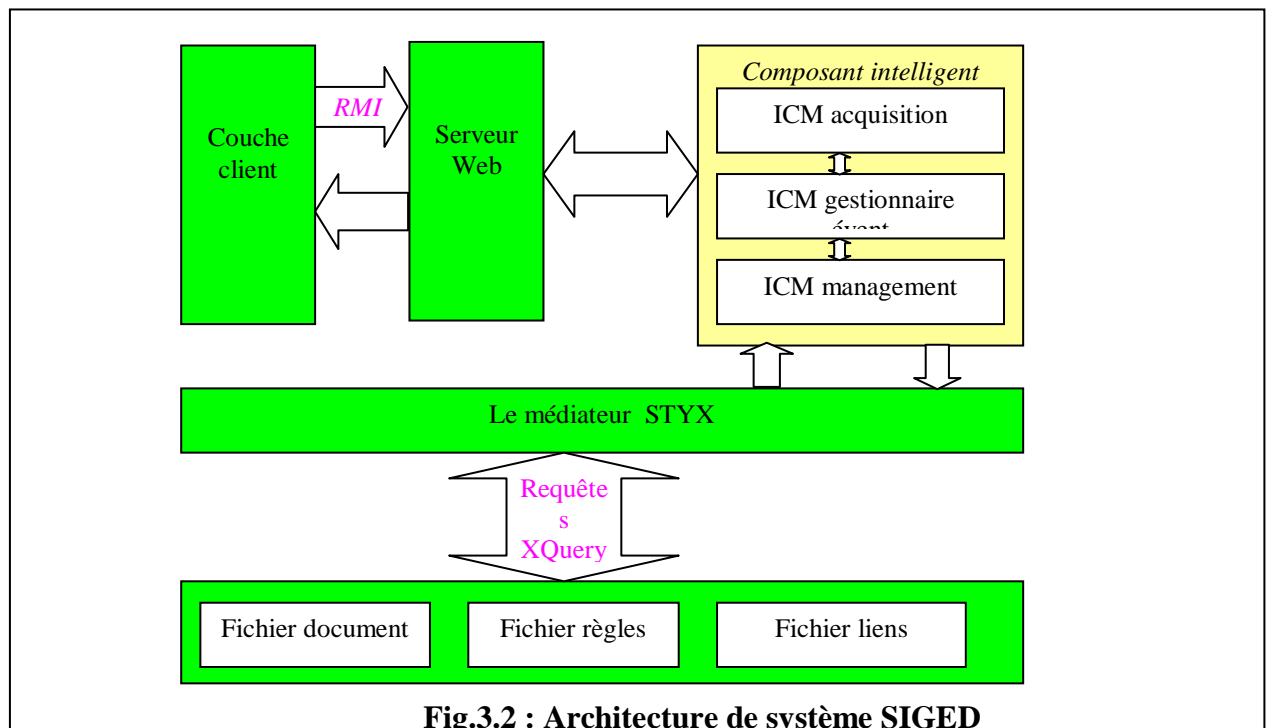

**Fig.3.2 : Architecture de système SIGED**

### a)   La couche application

C'est le module client proposé par le système à ses utilisateurs, il est doté d'une connexion au module serveur et ne s'adresse qu'à lui. Dans cette couche, l'utilisateur est considéré comme la partie interactive entre le serveur et ses composants.

Cette couche est considérée comme une interface homme machine qui permet à l'utilisateur d'accéder la base de données à travers des langages XML spécifiés. Les opérations sur la base peuvent être des opérations de mise à jour, des suppressions, des modifications des fragments ou des documents. Pour confirmer ces modifications un message instantané est produit.

Les requêtes envoyées par le client sont transférées au serveur actif qui gère les relations avec la base de données.

### b)   La couche serveur

Cette couche est responsable de la communication entre le client et la base de données, elle permet aussi de gérer les relations avec plusieurs modules clients distants. Cette couche permet d'établir la structure de document à partir de base de données, elle a deux interfaces de connexion, l'une avec le client en utilisant une interface RMI et l'autre  avec la base XML à travers un médiateur XML (STYX) [ABFS02].

Cette couche est composée de trois parties principales : le serveur Web, le composant intelligent management (ICM) et le médiateur STYX

**Description du « intelligent composant management » (ICM)**

Intelligent composant management (ICM): est un ensemble d'application extensible qui permet d'acquérir, de gérer et diffuser un contenu bien déterminé.
En effet, ce composant est considéré comme une suite logicielle qui est utilisée pour améliorer la gestion des documents et pour une meilleure gestion des règles actives.

Le composant « ICM » est composé de trois parties à savoir :
*Le composant « ICM acquisition » :* pour importer une partie d'un document XML (dans notre cas le contenu c'est la règle active),
*Le composant « ICM gestionnaire événement » :* est utilisé pour déclencher des événements relatifs à la lecture de la DTD de document XML, il est responsable aussi de la gestion de tous les événements qui ont été définis. Il permet de fournir les différentes méthodes qui sont associées à chaque événement détecté.
*Le composant « ICM management » :* permet d'exécuter les règles actives en utilisant des langages XML prédéfinis comme XQuery, XLink etc.
Ce composant intelligent permet l'extraction et l'exécution des règles intégrées dans un fichier XML envoyé par un utilisateur .Le processus de l'exécution des règles actives par ce composant est décrit dans la section suivante.

**c) La couche du stockage**

Les documents, les liens et les règles sont tous stockés dans des fichiers XML. Donc nous obtenons ainsi une orientation tout XML, ce qui rend la base homogène.
L'accès a la base XML native se fait à travers des langages purement XML ce qui permet d'offrir plus d'efficacité aux opérations de modifications de la base.

**3.3.4    Processus de l'exécution de la règle dans le système intelligent SIGED**

On décrit ci dessous le processus de l'exécution de la règle active par les différents sous composants du composant « intelligent composant management » (ICM).

**Le composant « ICM Acquisition  règle » :**

Ce composant  permet d'importer du contenu créé (fichier XML) une partie bien déterminée (la règle active) en utilisant le langage XPath. Ce langage permet de désigner une portion d'un document XML. À l'aide de ce composant, on peut importer n'importe quel format de document et à partir de diverses sources de données en utilisant plusieurs outils comme l'analyseur et des langages comme le XQuery    en    utilisant    la    technique    (for,    let,    where,    return)    [Rona02][Anke02].
Donc, a l'aide de XPath on peut repérer la règle contenue dans le fichier XML envoyé par le client  dans un premier temps, ensuite on extrait cette règle du fichier grâce au langage XQuery
En effet ICM acquisition assure deux  fonctions principales :
• Repérer la règle active  à partir de fichier  client en utilisant des langages XML comme XPath.
• Extraire la règle de fichier XML en utilisant XQuery.

**Le composant  « ICM gestionnaire événement » :**

Ce composant est utilisé pour déclencher des événements relatifs à la lecture de la DTD de document XML (c a d la règle : puisqu'elle est présentée sous forme d'un fichier XML).
Le gestionnaire d'événements, permet au composant « ICM  management » d'effectuer des  opérations selon le type d'élément rencontré.
Il est responsable de la gestion de tous les événements qui ont été définis. Il permet de fournir les différentes méthodes ou règles qui sont  associées à chaque événement défini.
L'événement est détecté par « le gestionnaire d'événement » à la suite de l'analyse de DTD correspondante à la règle active.
Ce composant permet la sélection des règles correspondantes à l'événement détecté en accédant au fichier règle à l'aide de langage XLink qui lui permet de pointer sur ce fichier et le langage XQuery pour extraire les règles  relatives à l'événement détecté. En effet le composant « ICM gestionnaire événement » peut accéder au fichier des règles XML pour repérer les règles qui correspondent à l'événement détecté en utilisant les langages XML comme XLink et XQuery, et par l'intermédiaire de médiateur STYX qui permet, à l'aide des règles de traduction et un algorithme de réécriture de requêtes, de traduire les requêtes de la part de composant intelligent en des requêtes XPath [ClDe99] ou XQuery [ChFR01].

***Le composant « ICM management règle » :***

        Il est utilisé pour exécuter les règles actives et organiser les actions à exécuter sur la base des documents. Il constitue le coeur de composant intelligent puisqu'il assure l'exécution des règles actives. C'est au niveau de ce composant que la règle active est exécutée.

Une fois les règles concernées par l'événement détecté sont sélectionnées par « ICM gestionnaire événement », le « ICM management » en utilisant le parseur valide, peut analyser toutes les règles qui correspondent à l'événement déclencheur détecté par le « ICM gestionnaire d'événement »

Le processus de l'exécution des actions des règles est décrit ci-dessous :

Le parseur XML permet de créer des structures des documents XML concernés par les changements et insère les DTD des actions des règles dans les templates rules (règles de gabarit).

Les templates rules sont des balises XSL permettant de définir des opérations à réaliser sur les éléments du document XML.

Le processeur XSLT (extensible stylesheet transformation) permet la génération de la structure logique du document XML et l'a fait subir des transformations en appliquant les actions insérées dans les templates rules.

Les DTD des actions des règles gardées sont insérées dans un document XSL a l'intérieur d 'une balise appelée < xsl : template >

Ce fichier XSL peut être lié au document XML par un URL (Href) en utilisant une balise comme la suivante :

**Exemple :**
```
< ? Xml  version ='1.0' encoding= 'ISO-8859-1" ?>
< ? Xml-stylesheet Href= »fichier.xsl' type ='text/xsl' ?>
```

On utilise les attributs fournis par les templates rules pour effectuer les changements apportés par les règles. L'attribut « match »de la balise <xsl : templates>permet de définir grâce de la notation Xpath les éléments a modifier.

**Exemple :**
<xsl : template match='personne'> permet de modifier l'élément de type personne.

La notation Xpath permet de définir des patterns pour repérer les éléments a transformés.

Donc, les actions sont insérées dans les balises <xsl : templates> .A l'aide des attributs match utilisant le langage Xpath et a l'aide des éléments de transformations, on peut effectuer les traitements apportés par la règle

**Exemple des éléments de transformations :**

L'élément apply template : utilisé dans la balise <xsl : template/>permet d'appliquer les changements apportés par la règle.

L'élément value of : utilisé pour insérer le contenu de la règle

L'élément <xsl : template> : permet de sélectionner les éléments sur les quels on désire effectuer des changements

L'élément for each : permet de sélectionner des éléments enfants et leurs appliquer des transformations.

        Donc, en utilisant les règles de gabarit qui permettent de définir des traitements à effectuer sur le document XML apportés par les règles actives et en appliquant les différents éléments de transformations,nous pouvons effectuer les changements apportés par les actions des règle sélectionnées après la satisfaction des conditions.

        Dans la section suivante, nous décrivons le processus de l'exécution de la règle active en utilisant UML pour voir l'enchaînement de l'exécution de la règle active dans le système SIGED.

### 3.3.5    Processus de l'exécution de la règle  en UML

Nous décrivons dans la section suivante la description de l'exécution de la règle dans le système de gestion des documents «  SIGED ». Le processus de l'exécution de la règle peut être divisé en quatre étapes :

1- L'extraction de la règle du document XML.
2- La détection de l'événement et la maintenance des changements.
3- Déclenchements et sélection des règles correspondantes.
4- Exécution des actions des règles sélectionnées après l'évaluation des conditions.

#### 3.3.5.1 Diagramme de séquence

L'exécution de la règle active dans notre système « SIGED » passe par plusieurs étapes depuis l'extraction jusqu'a l'exécution. En effet l'exécution d'une règle dans le système intelligent de gestion des documents « SIGED » est décrite de la façon  suivante :

Le client envoie un fichier XML. La règle est contenue dans le fichier XML qui est capté par le serveur Web et envoyé au composant intelligent « ICM ».Ce composant a pour but d'analyser le fichier XML envoyé par  l'utilisateur afin d'en extraire la règle active.

Le service « ICM acquisition » extrait la règle introduite dans le fichier XML à l'aide des langages  XML comme  XPointer et XQuery.

La règle est par la suite diffusée au composant « ICM gestionnaire d'évènement» pour détecter l'événement et déclencher les méthodes ou les règles relatives à cet  événement. Les règles sélectionnées sont envoyées au composant «ICM management » , ce dernier exécute les règles  et retourne le résultat  au client (Voir fig. 3.4).

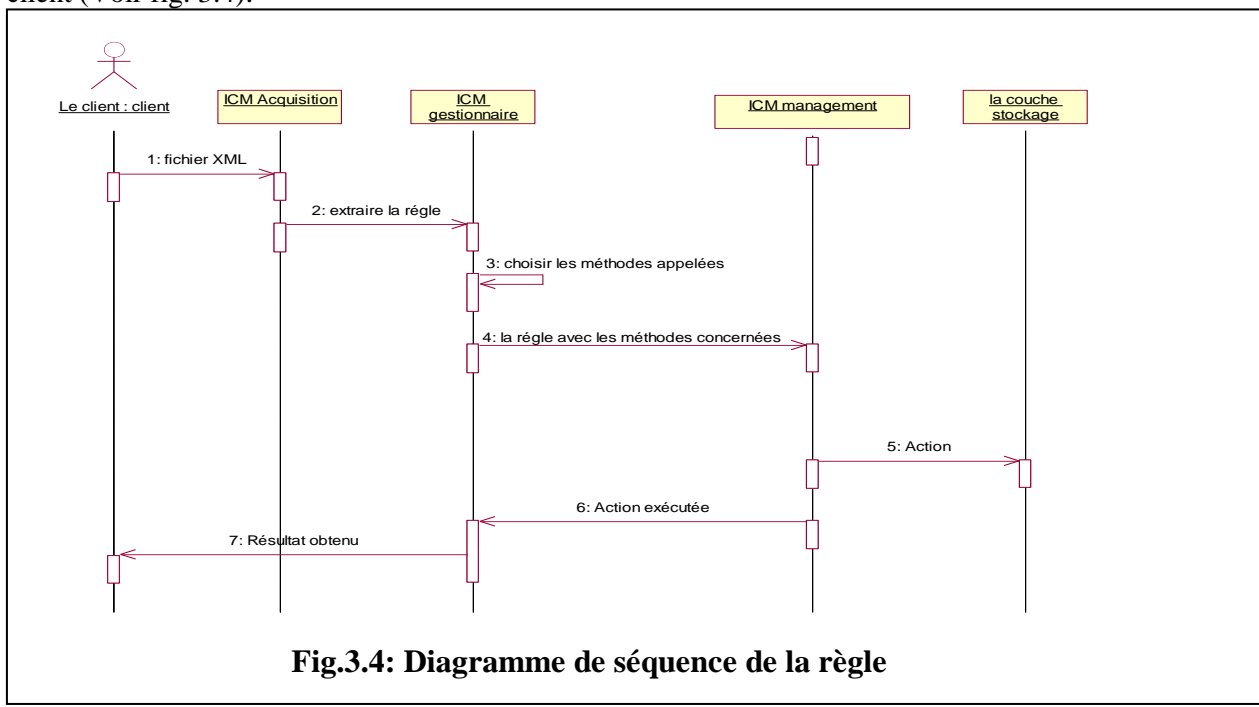

**Fig.3.4: Diagramme de séquence de la règle**

Nous décrivons dans la section suivante le model de l'exécution de la règle en utilisant le diagramme de classe UML pour voir l'interaction entre les différents éléments de la règle active.

#### 3.3.5.2 Le modèle  statique de l'exécution de la règle active

A l'aide de UML, nous décrivons les interactions entre les composants de la règle active.
 On représente le diagramme de classe pour l'exécution de la règle. En effet, l'événement est envoyé à travers une interface (RMI) par l'utilisateur permet le déclanchement de la règle active correspondante.
Le diagramme de classe du modèle d'exécution est présenté dans la figure ci-dessous. (Voir figure 3.5)

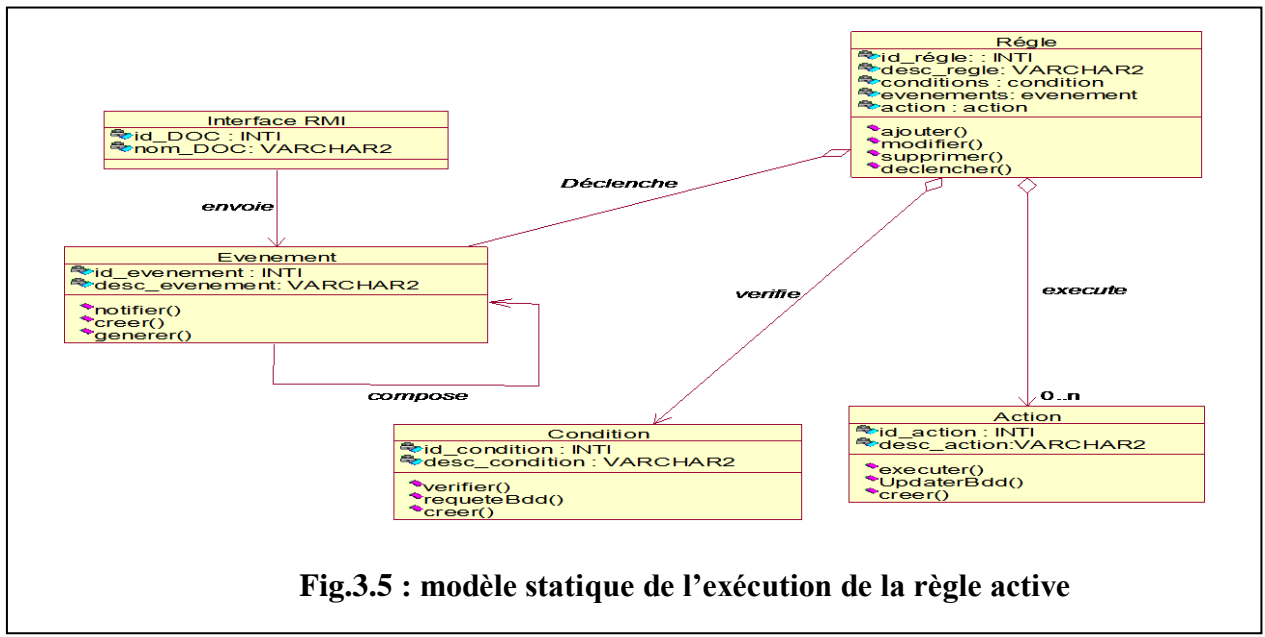

**Fig.3.5 : modèle statique de l'exécution de la règle active**

Nous présentons ci-dessous le diagramme de collaboration de l'exécution de la règle pour montrer les interactions entre les différents composants du système SIGED.

L'architecture de notre système SIGED est présentée sous forme d'une architecture client /serveur.

En effet, le client émet son fichier XML qui contient la règle à exécuter, le composant « ICM gestionnaire d'événement » déclenche l'événement et active le composant « ICM management » qui, à travers le Médiateur STYX, effectue les différents traitements apportés par la règle active

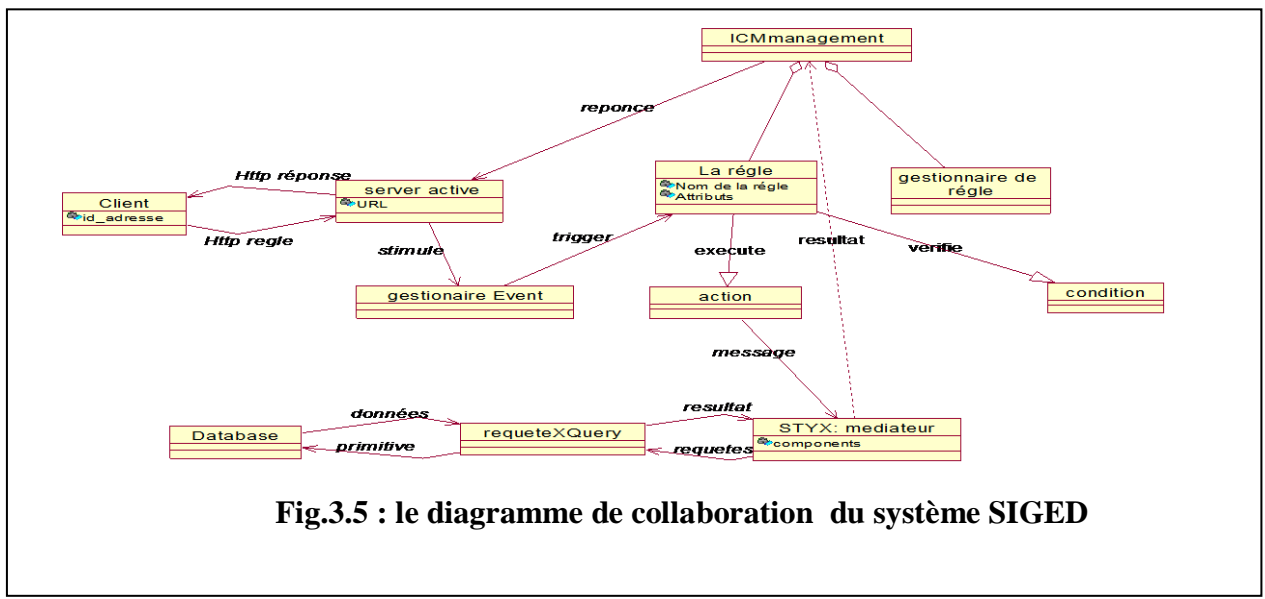

**Fig.3.5 : le diagramme de collaboration du système SIGED**

## 3.4     Le prototype de système SIGED

Afin d'illustrer notre système de gestion des documents « SIGED », nous réalisons ce système en utilisant le langage de programmation « Visual Studio 6.0 ».

Le client est connectée à l'application grâce à une interface crée en VB à travers la quelle l'utilisateur peut effectuer les différentes opérations de gestion des documents ou des règles.

En effet le programme doit permettre d'effectuer plusieurs taches : l'extraction des règles actives contenues dans un document XML,l'exécution des règles extraites des documents XML envoyés par l'utilisateur et finalement la gestion des règles actives ( l'ajout des règles , la suppression des règles, la modification des règles).

Suite à l'introduction d'une ou plusieurs règle(s) active(s) dans un fichier XML envoyé par l'utilisateur et à l'aide de commandes VB( *Selstart , Selenght* ), nous pouvons parcourir le document XML pour extraire la ou les règles actives qui se trouvent dans le fichier XML.

En effet, à l'aide de menu « **extraire** », on peut accéder au document XML qui contient les règles et afficher les règles avec les différents événements, conditions et actions de chacune de ces règles extraites (Voir annexe (Fig. 1)).

Après l'extraction, les noms des règles sont affichés dans un Combo Box (Voir annexe (Fig. 2)) avec ses différentes parties (événements, conditions, actions) (Voir annexe (Fig. 3)).

Pour l'exécution des règles actives, ce programme permet d'exécuter n'importe quel action ou opération apportée par les règles sélectionnées. (Voir annexe (Fig.4)).

A l'aide du menu « **gestion DOC XML** », nous pouvons effectuer à un document bien déterminé plusieurs opérations par exemple supprimer, ajouter ou modifier un élément du document XML. (Voir annexe (Fig.8)).

En plus ce programme permet la gestion des règles actives (ajout, suppression, modification)

Pour l'interface « **ajouter une règle** », nous pouvons ajouter une règle active dans un fichier XML règles. L'ajout se fait par l'ajout d'un ou plusieurs événements, une ou plusieurs conditions et une ou plusieurs action et par le bouton « *ajouter* » nous pouvons ajouter la règle formée dans le fichier de règles ou bien annuler l'opération d'ajout a l'aide de bouton « *annuler* » (Voir annexe(Fig.5)).

Pour l'interface « **supprimer une règle** », nous pouvons supprimer une règle active d'un fichier XML règle. La suppression se fait par le choix de la ou les règles concernées et par le bouton « *supprimer* » on peut supprimer la règle sélectionnée dans la liste ou bien annuler l'opération de suppression à l'aide de bouton « *annuler* » (Voir annexe(Fig.6)).

Pour l'interface « **modifier une règle** », nous pouvons modifier une règle active d'un fichier XML règle. La modification se fait par le choix de la ou les règle concernées et par le bouton « *modifier* » on peut modifier la règle sélectionnée dans la liste pour l'ajout, par suppression, par modification des différents éléments de la règle active (événement, condition, action) ou bien annuler l'opération de modification a l'aide de bouton « *annuler* » (Voir annexe (Fig.7)).

Concernant la gestion des documents XML, nous avons limité les opérations qui peuvent être exécuté dans un document XML entre l'ajout d'un élément et la suppression d'un élément.

Pour la suppression d'un élément dans un document XML (Voir annexe8), à l'aide de menu, « *supprimer élément* », nous cherchons l'élément à supprimer puis par le bouton « *supprimer* » nous confirmons la suppression de l'élément spécifié (Voir annexe(Fig.9)).

L'ajout d'un élément dans un document XML se fait par la spécification de la balise à ajouter et par la description de la balise (le texte à écrire entre la balise fermante et ouvrante). Par exemple, la balise à ajouté est « mémoire » et la description est « gestion des documents XML », donc l'élément ajouté dans le document XML est « *<mémoire> gestion des documents XML</mémoire>* » (Voir annexe (Fig.10)).

Ce programme permet gestion des documents XML actif contenant de paradigmes actifs (règles ECA) en permettant l'exécution de plusieurs actions, apportés par les règles actives, sur un document. En plus, ce programme permet la gestion des règles actives que ce soit par ajout, suppression ou modification des règles

## 3.5    Conclusion

Dans ce chapitre, nous avons présenté tout d'abord une modélisation de la couche de stockage de système « SIGED ». C'est une couche orientée tout XML obtenue par modélisation des bases de documents, des bases de règles et des bases de liens. Les documents, les règles et les liens sont représentés sous forme des DTD et sont stockés dans des fichiers XML.

Une base XML native est utilisée pour stocker les documents, les liens et les règles pour profiter des différents avantages apportés par ce type de base pour la recherche, l'indexation, l'intégrité référentielle etc.

La deuxième partie de ce chapitre est consacrée pour la présentation de notre système de gestion des documents « SIGED ».

Nous avons présentée l'architecture, les différents composants et les fonctionnalités de ce système. Nous avons présenté aussi description du processus de l'exécution de la règle active au niveau du système « SIGED », ce processus d'exécution est assuré par un composant intelligent nomme « ICM », utilisant des langages XML prédéfinies, permettant l'extraction et l'exécution des règles contenues au début dans un fichier XML envoyé par l'utilisateur.

Concernant l'implémentation, ce système permet de réaliser trois taches principales :

- L'extraction des règles actives à partir d'un document XML en utilisant des commandes VB pour l'extraction des règles (SelStart, Selenght)
- La gestion des documents XML par ajout ou suppression des éléments d'un document XML.
- La gestion des fichiers des règles que ce soit par ajout, suppression ou modification des règles.

Ce système peut être améliorer en prenons en compte d'autres types d'événements, l'intégration des fichiers des règles dans les documents XML etc.

# Conclusion générale

Nous avons abordé dans ce master les problèmes liés à la gestion des documents. En effet, les systèmes classiques pour la gestion de la documentation posent des problèmes puisqu'ils reposent sur des modèles des données propriétaires empêchant ainsi une bonne gestion des documents.

Notre contribution se base sur une approche visant à améliorer la gestion des documents. Dans ce contexte, nous avons présenté d'une part la définition d'une modélisation des différentes sources de données de la couche de stockage du système SAGED et nous avons proposé une couche de stockage orienté tout XML. En effet pour modéliser la couche de stockage nous avons modélisé ses différentes parties (la base des liens, la base des documents, la base des règles) que nous avons les présentés sous forme de DTD

Dans notre approche les liens , les documents et les règles sont stockés dans des fichiers XML ,le tout est stocké dans une base XML native qui offre plusieurs fonctionnalités pour la gestion des documents XML comme l'indexation , l'intégrité référentielle etc.

D'autre part, nous avons défini un système intelligent de gestion électronique de documents « SIGED » avec une architecture client /serveur, basée sur un composant intelligent permettant l'exécution des règles actives contenues dans un document XML.

En outre ce système permet l'exécution des règles actives en se basant sur un composant intelligent « composant ICM ». En effet, « composant ICM » est composé de trois sous composants, à savoir *« ICM acquisition »* pour extraire les règles actives à travers des langages XML prédéfinis comme XQuery et XPath, *« ICM gestionnaire d'événement »* pour détecter les événements de changement de la base afin d'extraire les règles concernées par l'événement détecté et *« ICM management »* pour exécuter les actions des règles actives en utilisant des outils XML prédéfinis comme le parseur XML et des langages comme XPath ,XSL et XSLT.

Les trois sous composants coopèrent entre eux pour extraire et exécuter les règles actives contenues au début dans un fichier XML envoyé par un utilisateur.

Nous avons décrit aussi le processus de l'exécution de la règle active en UML en le présentant par un diagramme de séquence et un diagramme de collaboration.

Le prototype de notre système est en cours de réalisation en « Visual Basic ». En effet, ce système permet de réaliser trois taches principales. D'abord, ce système permet l'extraction des règles actives à partir d'un document XML en utilisant des commandes VB pour l'extraction des règles (SelStart, SelLenght), ensuite, permet la gestion des documents XML par ajout ou suppression des éléments d'un document XML et enfin, permet la gestion des fichiers des règles que ce soit par ajout, suppression ou modification des règles.

Il est prévu également d'augmenter les possibilités de réaction de notre système en prenant en charge tous les événements possibles. Nous pouvons définir des mécanismes pour permettre la répartition des documents sur plusieurs serveurs, d'où la nécessité d'un mécanisme permettant les interactions entre ces différents serveurs. Aussi, nous pouvons augmenter notre système en intégrant les documents et les bases dans un seul fichier XML dans la base XML native.

# Références Bibliographiques

# ANNEXES

*Fig.1 :*

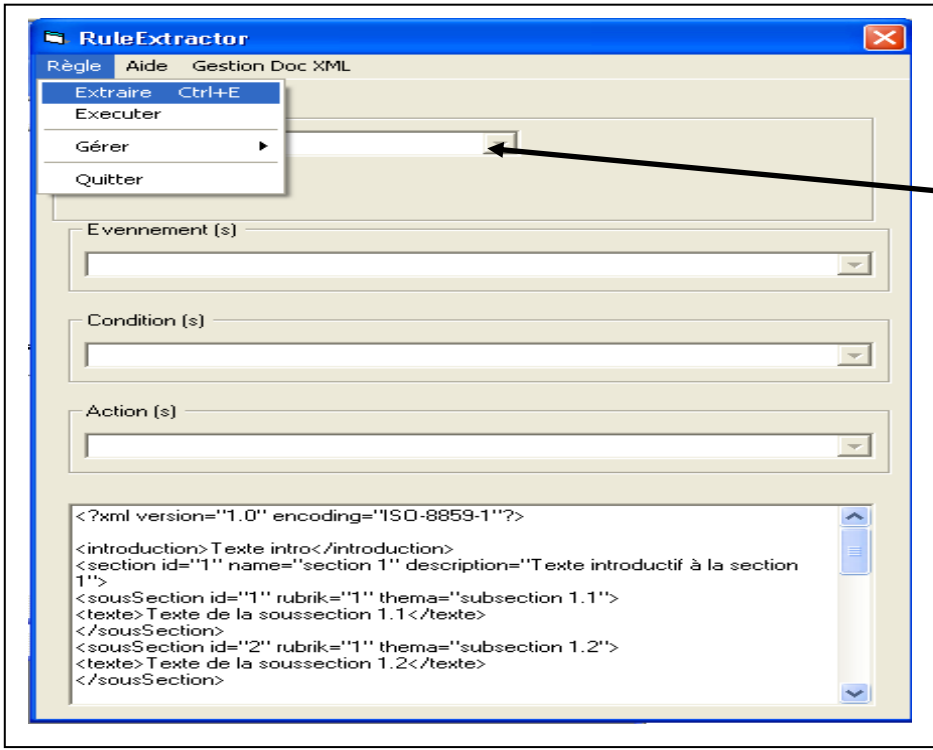

**Interface de l'application :**

- A l'aide de Menu *« Extraire »*, on peut extraire les règles du document XML.

  Les noms des règles sont affichés dans cette Combo Box.

- Les règles extraites sont affichées sous forme des événements, des conditions et des actions.

*Fig.2 :*

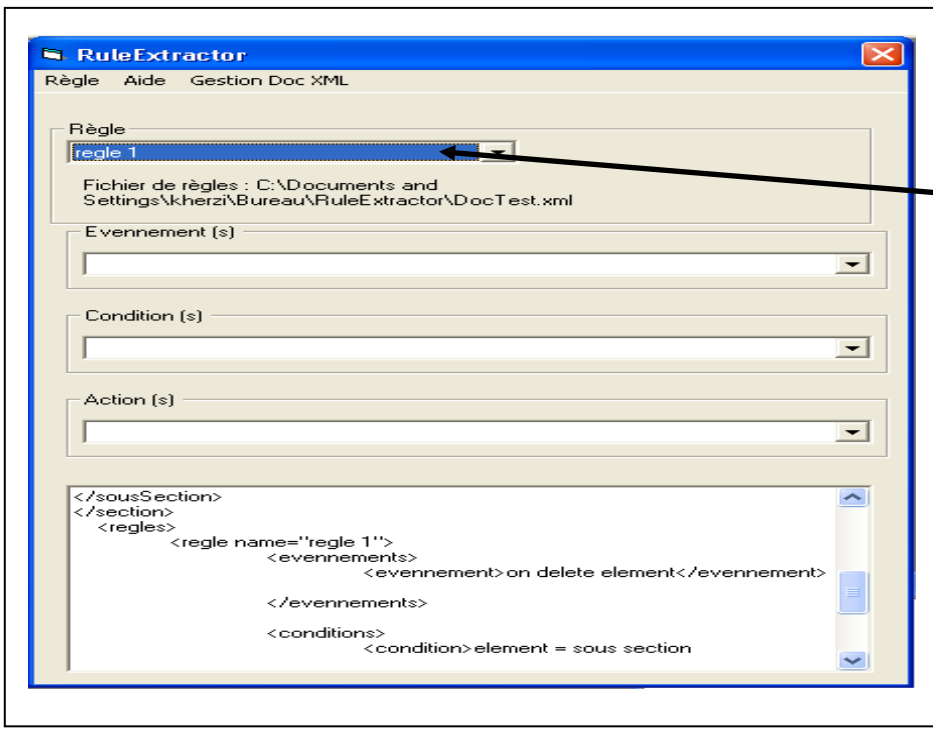

**Interface de l'extraction des règles :**
- Le document XML dans le TextBox, contenant une règle active (ECA) nommée règle 1.

  La règle nommée : règle 1 est extraite du document XML et affichée dans le Combo Box.

**Fig.3 :**

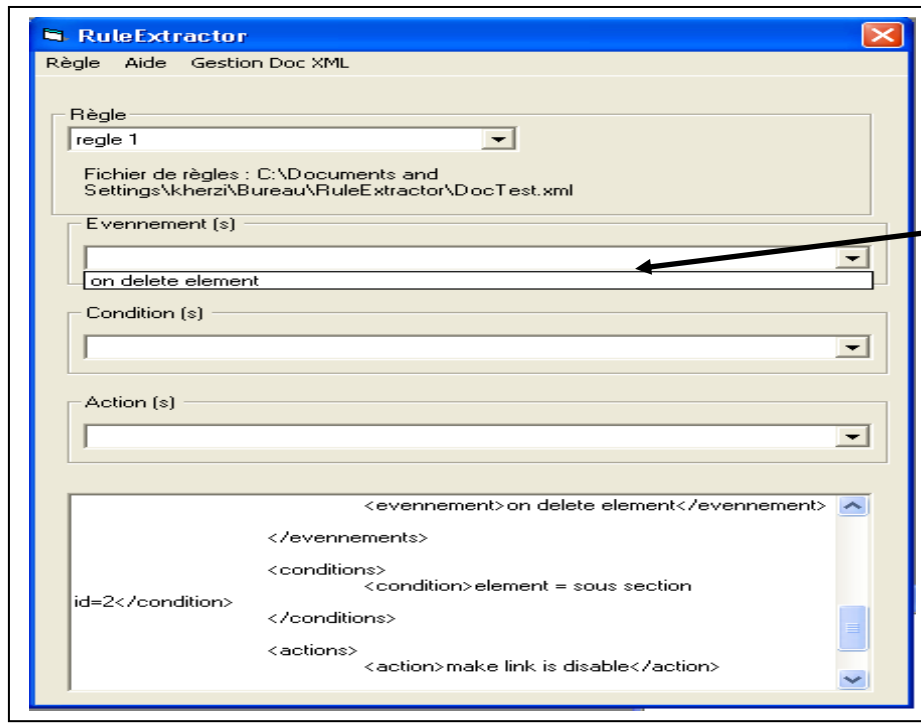

**Fig.4 :**

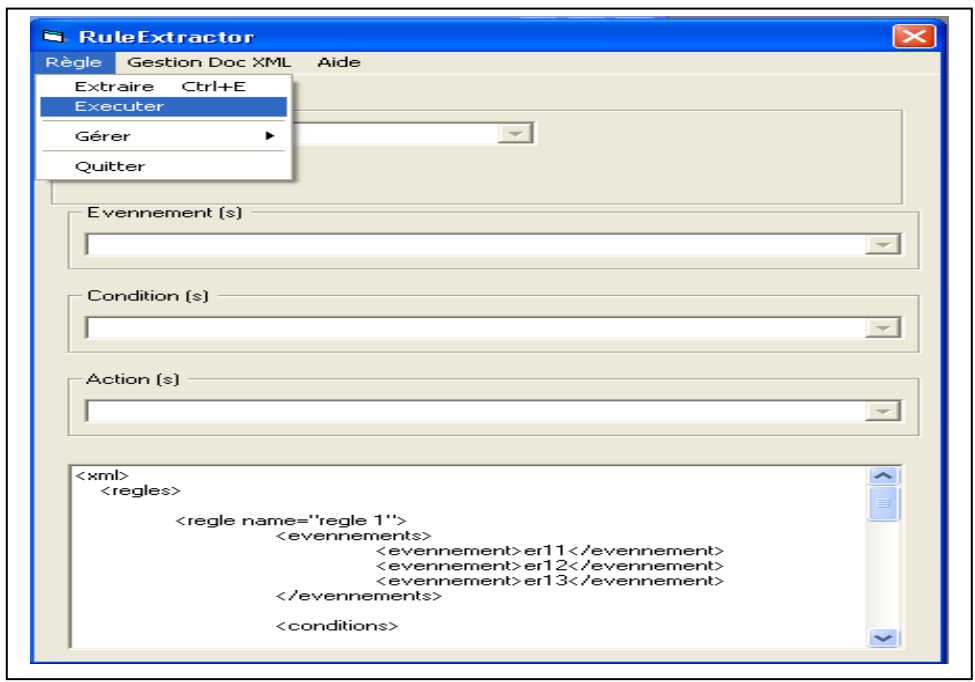

*Fig.5 :*

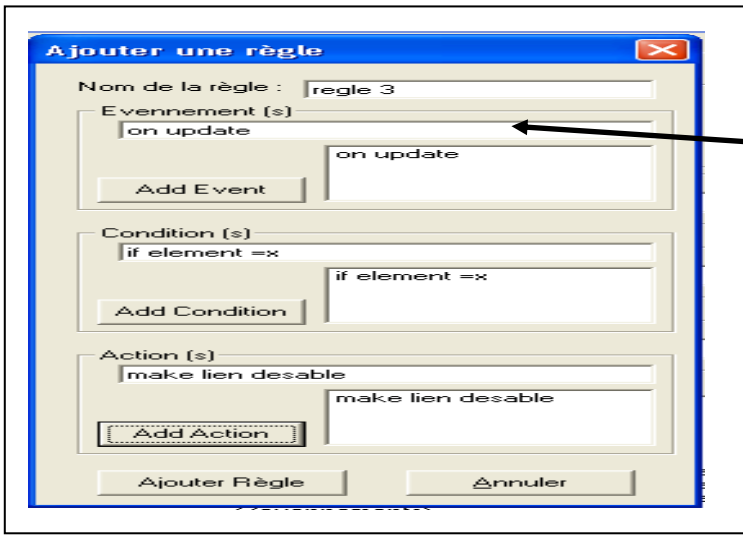

*Interface d'ajout d'une règle :*

- Pour ajouter la règle 3, on ajoute un ou plusieurs événements, condition, actions.
- On spécifie l'évènement à ajouter dans la zone de texte associée (on update)
- A l'aide de bouton *« Add évent »* on ajoute l'événement spécifié.
- Même chose pour les conditions et les actions.
- A l'aide du bouton *« ajouter règle »*, on ajoute cette règle au fichier règle

*Fig.6 :*

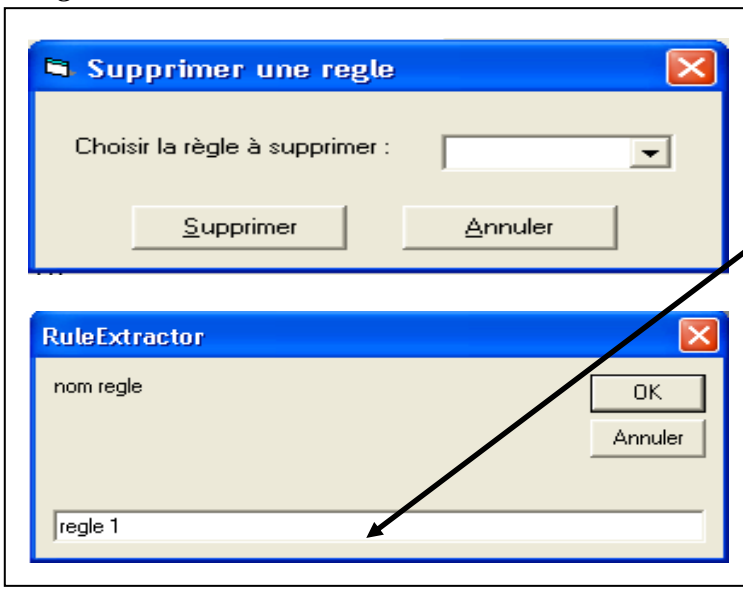

*Interface de suppression des règles :*

- La suppression des règles actives de fichier règle ou du document XML.
- Il suffit de spécifier le non de la règle à supprimée dans la zone de texte spécifié.
- A l'aide du bouton *« OK »*, on supprime la règle sélectionnée avec ses différentes parties (événements, conditions, actions)

*Fig.7 :*

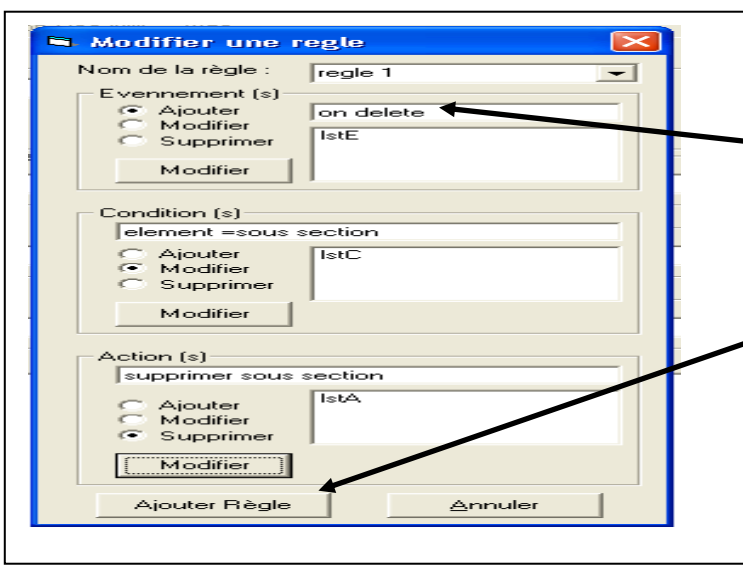

*Interface de modification des règles :*

- On récupère la règle a modifier avec ses différentes parties
- A chaque événement, on peut lui ajouter d'autres événements, le supprimer complètent ou la modifier par un autre
- Même traitement pour les conditions et les actions
- Le bouton *« ajouter règle »* permet d'ajouter la règle modifiée au document XML ou au fichier règle.

**Fig.8 :**

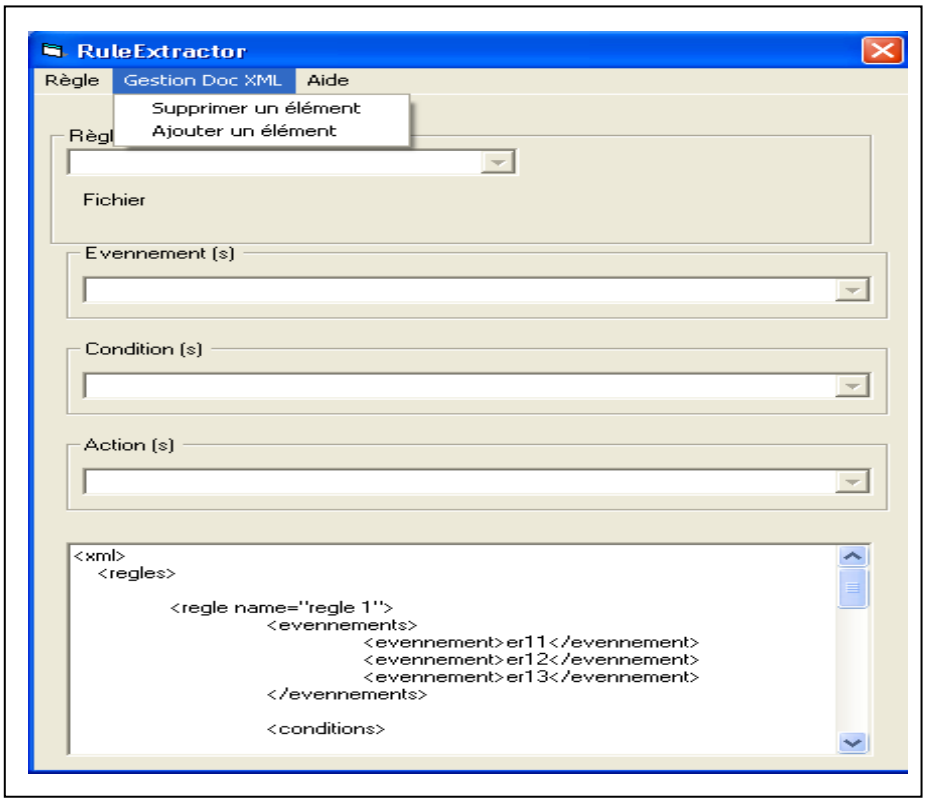

*Interface pour la gestion des documents XML :*

- On se limite pour la gestion des documents XML par ajout ou suppression d'un élément d'un document XML.
- Supprimer un élément d'un document XML revient a le rechercher puis le supprimer
- Ajouter un élément dans un document XML revient a le spécifier dans un zone de texte puis l'ajouter.

**Fig.9 :**

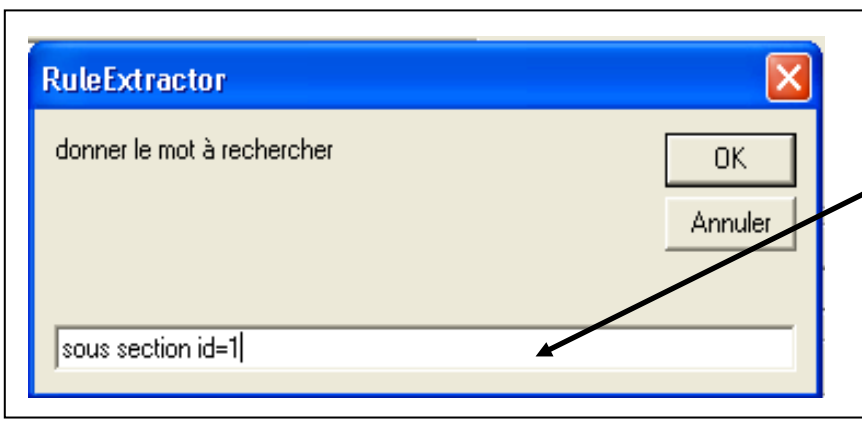

*Interface de suppression d'un élément :*

- Spécifier l'élément à supprimé
- A l'aide du bouton « OK », on supprime l'élément spécifié par parcours de document XML

**Fig.10 :**

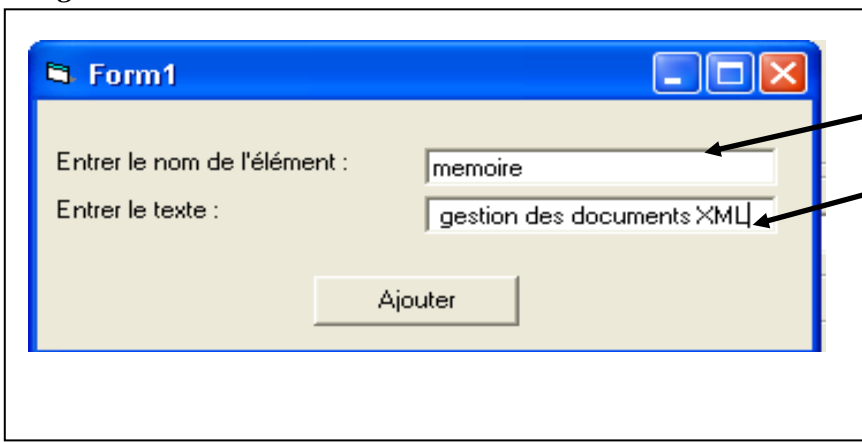

*Interface d'ajout d'un élément :*

- On entre le non de la balise a ajouter (mémoire)
- On spécifie le texte entre les deux balises (ouvrante et fermante)
- Le bouton *« ajouter »* permet d'ajouter la balise spécifiée dans le document XML
- L'élément a ajouté : <mémoire>la gestion des documents XML</mémoire>